\documentclass[journal,twocolumn,10pt]{IEEEtran}

\usepackage{graphicx}
\usepackage{amsfonts}
\usepackage{chngcntr}
\usepackage{times}
\usepackage{latexsym}
\usepackage{amssymb}
\usepackage{amsmath}
\usepackage{cite}
\usepackage{verbatim}
\usepackage{subfigure}
\usepackage{multirow}
\usepackage{lastpage}
\usepackage{calc}
\usepackage{eso-pic}
\usepackage{tabularx}
\usepackage{ifthen}
\usepackage{epstopdf}

\newtheorem{theorem}{Theorem}
\newcounter{as}
\newcounter{le}
\newcounter{cor}[theorem]

\newcounter{def}
\newcounter{pro}
\newcounter{cri}

\newtheorem{assumption}[as]{Assumption}

\newtheorem{corollary}[cor]{Corollary}
\counterwithin{cor}{theorem}
\newtheorem{criterion}[cri]{Criterion}
\newtheorem{definition}[def]{Definition}

\newtheorem{lemma}[le]{Lemma}

\newtheorem{proposition}[pro]{Proposition}

\newenvironment{proof}{ \textbf{Proof:} }{ \hfill $\Box$}

\newcommand{\figref}[1]{{Fig.}~\ref{#1}}

\def\bb0{{\mathbb{0}}}

\def\bb{{\mathbf{b}}}

\def\b0{{\mathbf{0}}}

\def\sf0{{\mathsf{0}}}

\usepackage{acronym}
\usepackage{url}
\acrodef{mmWave}{millimeter wave}
\acrodef{LOS}{line-of-sight}
\acrodef{NLOS}{non-line-of-sight}
\acrodef{SINR}{signal-to-noise-and-interference ratio}
\acrodef{RMS}{root mean square}
\acrodef{i.i.d.}{identically and independently distributed}
\acrodef{MIMO}{Multiple-input and multiple-output}
\acrodef{FBR}{front-to-back ratio}

\usepackage{footnote}
\allowdisplaybreaks

\begin{document}
\title{Coverage and Rate Analysis for \\Millimeter Wave Cellular Networks}

\author{
Tianyang Bai and Robert W. Heath, Jr.\\
\thanks{The authors are with The University of Texas at Austin, Austin, TX, USA.
(email: tybai@utexas.edu, rheath@utexas.edu) This work is supported in part by the National Science Foundation under Grants No. 1218338 and 1319556, and by a gift from Huawei Technologies, Inc.

 Preliminary results related to this paper were presented at the 1st IEEE Global Conference on Signal and Information Processing (GlobalSIP) \cite{bai2013b} and the $47$th Annual Asilomar Conference on Signals, Systems, and Computers (Asilomar) \cite{bai2013e}. }}

\maketitle
\begin{abstract}
Millimeter wave (mmWave) holds promise as a carrier frequency for fifth generation cellular networks. Because  mmWave signals are  sensitive to blockage,  prior models for cellular networks operated in the ultra high frequency (UHF)  band do not apply to analyze mmWave cellular networks directly. Leveraging concepts from stochastic geometry, this paper proposes a general framework to evaluate the coverage and rate performance in mmWave cellular networks. Using a distance-dependent line-of-site (LOS) probability function, the locations of the LOS and non-LOS base stations are modeled as two independent non-homogeneous Poisson point processes, to which different path loss laws are applied. Based on the proposed framework, expressions for the signal-to-noise-and-interference ratio (SINR) and rate coverage probability are derived. The mmWave coverage and rate performance are examined as a function of the antenna geometry and base station density. The case of dense networks is further analyzed by applying a simplified system model, in which the LOS region of a user is approximated as a fixed LOS ball. The  results show that dense mmWave networks can achieve comparable coverage and much higher data rates than conventional UHF cellular systems, despite the presence of blockages. The results suggest that the cell size to achieve the optimal SINR scales with the average size of the area that is LOS to a user.
\end{abstract}

\section{Introduction}

The large available bandwidth at \ac{mmWave} frequencies makes them attractive for fifth generation cellular networks \cite{rappaport2014,pi2011,Rappaport2013a}. The \ac{mmWave} band ranging from 30 GHz to 300 GHz has already been considered in various commercial wireless systems including IEEE 802.15.3c for personal area networking \cite{Baykas2011}, IEEE 802.11ad for local area networking \cite{802.11ad}, and IEEE 802.16.1 for fixed-point access links \cite{802.16}. Recent field measurements reveal the promise of \ac{mmWave} signals for the access link (between the mobile station and base station) in cellular systems  \cite{Rappaport2013,Rappaport2013a}.

One differentiating feature of \ac{mmWave} cellular communication is the use of antenna arrays at the transmitter and receiver to provide array gain. As the wavelength decreases, antenna sizes also decrease, reducing the antenna aperture. For example, from the Friis free-space equation \cite{Goldsmith2005}, a \ac{mmWave} signal at 30 GHz will experience 20 dB larger path loss than a signal at 3 GHz. Thanks to the small wavelength, however, it is possible to pack multiple antenna elements into the limited space at \ac{mmWave} transceivers \cite{rappaport2014}. With large antenna arrays, \ac{mmWave} cellular systems can implement beamforming at the transmitter and receiver to provide array gain that compensates for the frequency-dependent path loss, overcomes additional noise power, and as a bonus also  reduces out-of-cell interference \cite{pi2011}.

Another distinguishing feature of \ac{mmWave} cellular communication is the propagation environment. MmWave signals are more sensitive to blockage effects than signals in lower-frequency bands, as certain materials like concrete walls found on building exteriors cause severe penetration loss \cite{alejos2008}. This indicates that indoor users are unlikely to be covered by outdoor \ac{mmWave} base stations. Channel measurements using directional antennas \cite{Rajagopal2012, Rappaport2013,Rappaport2013a} have revealed another interesting behavior at \ac{mmWave}: blockages cause substantial differences in the \ac{LOS}
paths and \ac{NLOS} path loss characteristics. Such differences have also been observed in prior propagation studies at ultra high frequency bands (UHF) from 300 MHz to 3 GHz, e.g. see \cite{3GPPTR36.8142010}. The differences, however, become more significant for \ac{mmWave} since diffraction effects are negligible \cite{pi2011}, and there are only a few scattering clusters \cite{akdeniz2013a}. Measurements in \cite{Rajagopal2012,Rappaport2013,Rappaport2013a} showed that mmWave signals propagate as in free space with a path loss exponent of 2. The situation was different for \ac{NLOS} paths where a log distance model was fit with a higher path loss exponent and additional shadowing \cite{Rappaport2013,Rappaport2013a}. The \ac{NLOS} path loss laws tend to be more dependent  on the scattering environment.
For example, an exponent as large as 5.76 was found in downtown New York City \cite{Rappaport2013a}, while only 3.86 was found on the UT Austin campus \cite{Rappaport2013}. The distinguishing features of the propagation environment need to be incorporated into the any comprehensive system analysis of \ac{mmWave} networks.

The performance of \ac{mmWave} cellular networks was simulated in prior work \cite{akdeniz2013a,akdeniz2013} using insights from propagation channel measurements \cite{Rappaport2013a}.
 In \cite{akdeniz2013}, using the \ac{NLOS} path loss law measured in the New York City,   lower bounds of the \ac{SINR} distribution and the achievable rate were simulated in a 28 GHz pico-cellular system. In \cite{akdeniz2013a}, a mmWave channel model that incorporated blockage effects and angle spread was proposed and further applied to simulate the mmWave network capacity. Both results in \cite{akdeniz2013a, akdeniz2013} show that the achievable rate in mmWave networks can outperform conventional cellular networks in the ultra high frequency (UHF) band by an order-of-magnitude. The simulation-based approach \cite{akdeniz2013,akdeniz2013a} does not lead to elegant system analysis as in \cite{Andrews2011a}, which can be broadly applied to different deployment scenarios.

Stochastic geometry is a useful tool to analyze system performance in conventional cellular networks \cite{Andrews2011a,Ghosh2012}.
In \cite{Andrews2011a}, by modeling base station locations in a conventional cellular network as a Poisson point process (PPP) on the plane, the aggregate coverage probability was derived in a simple form, e.g. a closed-form expression when the path loss exponent is 4. Moreover, the stochastic model was shown to provide a lower bound of the performance in a real cellular system \cite{Andrews2011a}.
There have been several extensions of the results in \cite{Andrews2011a}, such as analyzing a multi-tier network in \cite{Dhillon2012} and predicting the site-specific performance in heterogeneous networks in \cite{Heath2012}. It is not possible to directly apply results from conventional networks to \ac{mmWave} networks due to the different propagation characteristics and the use of directional beamforming. There has been limited application of stochastic geometry to study  \ac{mmWave} cellular networks. The primary related work was in \cite{akoum2012}, where directional beamforming was incorporated for single and multiple user configurations, but a simplified path loss model was used that did not take \ac{mmWave} propagation features into account.

A systematic study of \ac{mmWave} network performance should incorporate the impact of blockages such as buildings in urban areas. One approach is to model the blockages explicitly in terms of their sizes, locations, and shapes using data from a geographic information system. This approach is well suited for site-specific simulations \cite{Seidel1994} using electromagnetic simulation tools like ray tracing \cite{Toscano2003}. An alternative is to employ a stochastic blockage model, e.g. \cite{Franceschetti2004,bai2013}, where the blockage parameters are drawn randomly according to some distribution. The stochastic approach lends itself better to system analysis and can be applied to study system deployments under a variety of blockage parameters such as size and density.

The main contribution of this paper is to propose a stochastic geometry framework for analyzing the coverage and rate in \ac{mmWave} cellular networks. As a byproduct, the framework also applies to analyze heterogenous networks in which the base stations are distributed as certain non-homogeneous PPPs. 
We incorporate directional beamforming by modeling the beamforming gains as marks of the base station PPPs. For  tractability of the analysis, the actual beamforming patterns are also approximated by a sectored model, which characterizes key features of an antenna pattern: directivity gain, half-power beamwidth, and front-back ratio. A similar model was also employed in work on ad hoc networks \cite{Hunter2008}.
To incorporate blockage effects, we model the probability that a communication link is \ac{LOS} as a function of the link length, and provide a stochastic characterization of the region where a user does not experience any blockage, which we define as the {\it LOS region}. Applying the distance-dependent \ac{LOS} probability function, the base stations are equivalently divided into two independent non-homogenous point processes on the plane: the \ac{LOS} and the \ac{NLOS} base station processes. Different path loss laws and fading are applied separably to the \ac{LOS} and \ac{NLOS} case.  Based on the system model, expressions for the \ac{SINR} and rate coverage probability are derived in general mmWave networks. To simplify the analysis, we also propose a systematic approach to approximate a complicated LOS function as its equivalent step function. Our analysis indicates that the coverage and rate are sensitive to the density of base stations and the distribution of blockages in mmWave networks. It also shows that dense mmWave networks can generally achieve good coverage and significantly higher achievable rate than conventional cellular networks.

 A simplified system model is proposed to analyze dense mmWave networks, where the infrastructure density is comparable to the blockage density. For a general \ac{LOS} function, the LOS region observed by a user has an irregular and random shape. Coverage analysis requires integrating the SINR over this region \cite{bai2013b}. We propose to simplify the analysis by approximating the actual \ac{LOS} region as a fixed-sized ball called the {\it equivalent LOS ball}.  The radius of the equivalent LOS ball is chosen so that the ball has the same average number of \ac{LOS} base stations in the network.  With the simplified network model, we find that in a dense mmWave network, the cell radius should scale with the size of \ac{LOS} region to maintain the same coverage probability. We find that continuing to increase base station density (leading to what we call ultra-dense networks) does not always improve SINR, and the optimal base station density should be finite.


Compared with our prior work in \cite{bai2013b}, this paper provides a generalized mathematical framework and includes the detailed mathematical derivations. The system model applies for a general \ac{LOS} probability function and includes the impact of general small-scale fading. We also provide a new approach to compute coverage probability, which avoids inverting the Fourier transform numerically and is more efficient than prior expressions in \cite{bai2013b}. Compared with our prior work in \cite{bai2013e}, we also remove the constraint that the \ac{LOS} path loss exponent is 2, and extend the results in \cite{bai2013e} to general path loss exponents, in addition to providing derivations for all results, and new simulation results.

This paper is organized as follows. We introduce the system model in Section \ref{sec:system}. We derive  expressions for the \ac{SINR} and rate coverage in a general mmWave network in Section \ref{sec:generalnetwork}. A systematic approach is also proposed to approximate general \ac{LOS} probability functions as a step function to further simplify analysis. In Section \ref{sec:dense}, we apply the simplified system model to analyze performance and examine asymptotic trends in dense mmWave networks, where outdoor users observe more than one \ac{LOS} base stations with high probability. Finally, conclusions and suggestions for future work are provided in Section \ref{sec:conclusion}.

\section{System Model}\label{sec:system}
\begin{figure}[!ht]
\centering
\subfigure[center][{System model for mmWave cellular networks}]{
\includegraphics[width=0.8\columnwidth]{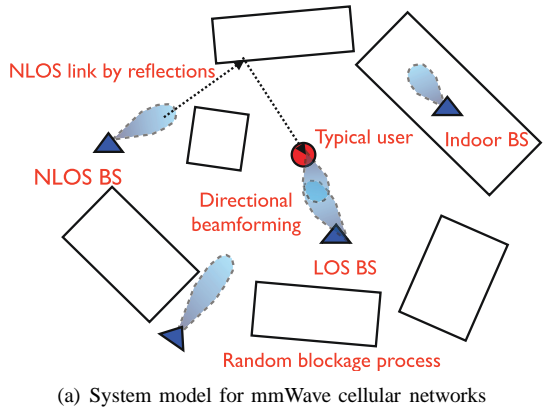}\label{fig:system_model}}
\subfigure[center][{Sectored model to approximate beamforming patterns.}]{
\includegraphics[width=0.8\columnwidth]{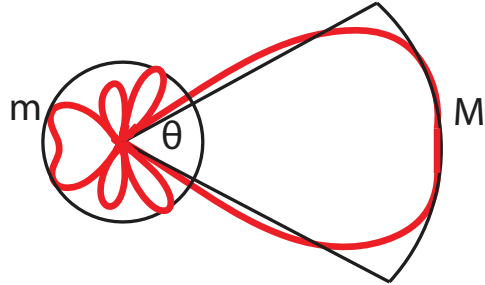}\label{fig:antenna}}
\caption{ In (a), we illustrate the proposed system model for mmWave cellular networks. Blockages are modeled as a random process of rectangles, while base stations are assumed to be distributed as a Poisson point process on the plane. An outdoor typical user is fixed at the origin. The base stations are categorized into three groups: indoor base stations, outdoor base stations that are \ac{LOS} to the typical user, and outdoor base stations that are \ac{NLOS} to the user. Directional beamforming is performed at both base stations and mobile stations to exploit directivity gains. In (b), we illustrate the sectored antenna model $G_{M,m,\theta}(\phi)$, which is used to approximate the beamforming patterns.}
\end{figure}


In this section, we introduce our system model for evaluating the performance of a \ac{mmWave} network. We focus on  downlink coverage and rate  experienced by an outdoor user, as illustrated in Fig. \ref{fig:system_model}. We make the following assumptions in our mathematical formulation.

\begin{assumption}[Blockage process]
The blockages, typically buildings in urban areas, form a process of random shapes, e.g. a Boolean scheme of rectangles \cite{bai2013}, on the plane. We assume the distribution of the blockage process to be stationary and isotropic - in other words - invariant to the motions of translation and rotation \cite[Chapter 10]{BacBla:Stochastic-Geometry-and-Wireless:09}.
\end{assumption}

\begin{assumption}[PPP BS]
The base stations form a homogeneous PPP $\tilde{\Phi}$ with density $\tilde{\lambda}$ on the plane. Note that a base station can be located either inside a blockage or outside a blockage. In this paper, however, we will focus on the SINR and rate provided by the outdoor base stations as the blockages are assumed to be impenetrable. Let $\Phi=\{X_\ell\}$ be the point process of outdoor base stations, $X_\ell$ the $\ell$-th outdoor base station, and $R_\ell=|OX_\ell|$ denote the distance from $\ell$-th base station to the origin $O$. Define $\tau$ as the average fraction of the land covered by blockages, i.e., the average fraction of indoor area in the network. Further, we assume the base station process $\tilde{\Phi}$ is independent of the blockage process. Therefore, each base station has an i.i.d. probability $1-\tau$ to be located outdoor. By the thinning theorem of PPP \cite{BacBla:Stochastic-Geometry-and-Wireless:09}, the outdoor base station process $\Phi$ is a PPP of density $\lambda=(1-\tau)\tilde{\lambda}$ on the plane. In addition, all base stations are assumed to have a constant transmit power $P_{\mathrm{t}}$.
\end{assumption}

\begin{assumption}[Outdoor user]
The users are distributed as a stationary point process independent of the base stations and blockages on the plane. A typical user is assumed to be located at the origin $O$, which is a standard approach in the analysis using stochastic geometry \cite{Andrews2011a,BacBla:Stochastic-Geometry-and-Wireless:09}. By the stationarity and independence of the user process, the downlink SINR and rate experienced by the typical user have the same distributions as the aggregate ones in the network.
The typical user is assumed to be outdoors. The indoor-to-outdoor penetration loss is assumed to be high enough such that an outdoor user can not receive any signal or interference from an indoor base station. Therefore, the focus in this paper is on investigating the conditional \ac{SINR} and rate distribution of the outdoor typical user served by outdoor infrastructure. Indoor users can be served by either indoor base stations or by outdoor base stations operated at UHF frequencies, which have smaller indoor-to-outdoor penetration losses in many common building materials. We defer the extension to incorporate indoor users to future work.
\end{assumption}

We say that a base station at $X$ is {\it\ac{LOS}} to the typical user at the origin $O$ if and only if there is no blockage intersecting the link $OX$. Due to the presence of blockages, only a subset of the outdoor base stations $\Phi$ are \ac{LOS} to the typical user.
\begin{assumption}[LOS and NLOS BS]\label{assump:LOS}
An outdoor base station can be either \ac{LOS} or \ac{NLOS} to the typical user. Let $\Phi_\mathrm{L}$ be the point process of \ac{LOS} base stations, and $\Phi_\mathrm{N}=\Phi/\Phi_\mathrm{L}$ be the process of \ac{NLOS} base stations. Define the {\it\ac{LOS} probability function} $p(R)$ as the probability that a link of length $R$ is \ac{LOS}. Noting the fact that the distribution of the blockage process is stationary and isotropic, the LOS probability function depends only on the length of the link $R$. Also, $p(R)$ is a non-increasing function of $R$; as the longer the link, the more likely it will be intersected by one or more blockages. The \ac{NLOS} probability of a link is $1-p(R)$.
\end{assumption}

The \ac{LOS} probability function in a network can be derived from field measurements \cite{akdeniz2013a} or stochastic blockage models \cite{bai2013,Franceschetti2004}, where the blockage parameters are characterized by some random distributions. For instance, when the blockages are modeled as a rectangle boolean scheme in \cite{bai2013}, it follows that $p(R)=\mathrm{e}^{-\beta R}$, where $\beta$ is a parameter determined by the density and the average size of the blockages, and $1/\beta$ is what we called the average \ac{LOS} range of the network in \cite{bai2013}.

For the tractability of analysis, we further make the following independent assumption on the LOS probability; taking account of the correlations in blockage effects generally makes the exact analysis difficult.
\begin{assumption}[Independent LOS probability]\label{assump:shadowing}
The LOS probabilities are assumed to be independent between different links, i.e., we ignore potential correlations of blockage effects between links.
\end{assumption}

Note that the \ac{LOS} probabilities for different links are not independent in reality. For instance, neighboring base stations might be blocked by a large building simultaneously. Numerical results in \cite{bai2013}, however, indicated that ignoring such correlations cause a minor loss of accuracy in the \ac{SINR} evaluation. Assumption \ref{assump:shadowing} also indicates that the \ac{LOS} base station process $\Phi_\mathrm{L}$ and the \ac{NLOS} process $\Phi_\mathrm{N}$ form two independent non-homogeneous PPP with the density functions $p(R)\lambda$ and $(1-p(R))\lambda$, respectively, where $R$ is the radius in polar coordinates.

\begin{assumption}[Path loss model]
Different path loss laws are applied to \ac{LOS} and \ac{NLOS} links. Given a link has length $R$, its path loss gain $L(R)$ 
is computed as
\begin{align}
L(R) = \mathbb{I}(p(R)) C_\mathrm{L} R^{-\alpha_{\mathrm{L}}}+  (1-\mathbb{I}(p(R)) C_\mathrm{N} R^{-\alpha_{\mathrm{N}}},
\end{align}
where $\mathbb{I}(x)$ is a Bernoulli random variable with parameter $x$, $\alpha_{\mathrm{L}}$, $\alpha_{\mathrm{N}}$ are the \ac{LOS} and \ac{NLOS} path loss exponents, and $C_\mathrm{L}$, $C_\mathrm{N}$ are the intercepts of the LOS and NLOS path loss formulas. Typical values of \ac{mmWave} path loss exponents and intercept constants are available in prior work, see e.g. \cite{Rappaport2013,Rappaport2013a}. The model could be further enhanced by including log-normal shadowing, but this is deferred in our paper to simplify the analysis.
\end{assumption}


\begin{assumption}[Directional beamforming]\label{assump:sector}
Antenna arrays are deployed at both base stations and mobile stations to perform directional beamforming. For tractability of the analysis, the actual array patterns are approximated by a sectored antenna model, which was used in prior ad hoc network analysis \cite{Hunter2008}. Let $G_{M,m,\theta}(\phi)$ denote the sectored antenna pattern in Fig. 1(b), where $M$ is the main lobe directivity gain, $m$ is the back lobe gain, $\theta$ is the beamwidth of the main lobe, and $\phi$ is the angle off the boresight direction. In the sectored antenna model, the array gains are assumed to be constant $M$ for all angles in the main lobe, and another constant $m$ in the side lobe in the sectored model.
We let $M_\mathrm{t}$, $m_\mathrm{t}$, and $\theta_\mathrm{t}$ be the main lobe gain, side lobe gain, and half power beamwidth of the base station antenna, and $M_\mathrm{r}$, $m_\mathrm{r}$, and $\theta_\mathrm{r}$ the corresponding parameters for the mobile station. Without loss of generality, we denote the boresight direction of the antennas as $0^\circ$. Further, let $D_\ell=G_{M_\mathrm{t},m_\mathrm{t},\theta_\mathrm{t}}(\phi_\mathrm{t}^\ell)G_{M_\mathrm{r},m_\mathrm{r},\theta_\mathrm{r}}(\phi_\mathrm{r}^\ell)$ be the total directivity gain in the link from the $\ell$-th base station to the typical user, where $\phi_\mathrm{r}^\ell$ and $\phi_\mathrm{t}^\ell$ are the angle of arrival and the angle of departure of the signal.

\end{assumption}


\begin{assumption}[User association]\label{assump:user}
The typical user is associated with the base station, either LOS or NLOS, that has the smallest path loss $L(R_\ell)$. The serving base station is denoted as $X_0$. Both the mobile station and its serving base station will estimate channels including angles of arrivals and fading, and then adjust their antenna steering orientations accordingly to exploit the maximum directivity gain. Errors in channel estimation are neglected, and so are errors in time and carrier frequency synchronizations in our work. Thus, the directivity gain for the desired signal link is $D_0=M_\mathrm{r}M_\mathrm{t}$. For the $\ell$-th interfering link, the angles $\phi^\ell_\mathrm{r}$ and $\phi^\ell_\mathrm{t}$ are assumed to be independently and uniformly distributed in $(0,2\pi]$, which gives a random directivity gain $D_\ell$.
\end{assumption}

By Assumption \ref{assump:sector} and Assumption \ref{assump:user}, the directivity gain in an interference link $D_\ell$ is a discrete random variable with the probability distribution as
$
D_\ell=a_k
$ with probability $b_k$ $(k\in\{1,2,3,4\})$, where $a_k$ and $b_k$ are constants defined in Table \ref{table:constant}, $c_\mathrm{r}=\frac{\theta_\mathrm{r}}{2\pi}$, and $c_\mathrm{t}=\frac{\theta_\mathrm{t}}{2\pi}$.
\begin{table}
\begin{center}
\caption{Probability Mass Function of $D_\ell$ and $\bar{D}_\ell$}\label{table:constant}
\small{\begin{tabular}{ c | c | c |c |c}
\hline
    \mbox{k} & 1&2&3&4 \\ \hline
    \mbox{$a_k$} &$M_\mathrm{r}M_\mathrm{t}$& $M_\mathrm{r}m_\mathrm{t}$  & $m_\mathrm{r}M_\mathrm{t}$&$m_\mathrm{r}m_\mathrm{t}$\\ \hline
    \mbox{$b_k$}&$c_\mathrm{r}c_\mathrm{t}$ &$c_\mathrm{r}(1-c_\mathrm{t})$  &$(1-c_\mathrm{r})c_\mathrm{t}$ &$(1-c_\mathrm{r})(1-c_\mathrm{t})$\\ \hline
\mbox{$e_k$} &$M_\mathrm{r}$& $M_\mathrm{r}/\xi_\mathrm{t}$  & $m_\mathrm{r}$&$m_\mathrm{r}/\xi_\mathrm{t}$\\ \hline
 \end{tabular}}
  \end{center}
\end{table}
\begin{assumption}[Small-scale fading]
We assume independent Nakagami fading for each link. Different parameters of Nakagami fading $N_\mathrm{L}$ and $N_\mathrm{N}$ are assumed for \ac{LOS} and \ac{NLOS} links. Let $h_\ell$ be the small-scale fading term on the $\ell$-th link. Then $\left|h_\ell \right|^2$ is a normalized Gamma random variable. Further, for simplicity, we assume $N_\mathrm{L}$ and $N_\mathrm{N}$ are positive integers.  We also ignore the frequency selectivity in fading, as measurements show that the delay spread is generally small \cite{Rappaport2013a}, and the impact of frequency-selective fading can be minimized by techniques like orthogonal frequency-division multiplexing or frequency domain equalization \cite{Goldsmith2005}.
\end{assumption}

Measurement results indicated that small-scale fading at \ac{mmWave} is less severe than that in conventional systems when narrow beam antennas are used \cite{Rappaport2013a}. Thus, we can use a large Nakagami parameter $N_\mathrm{L}$ to approximate the small-variance fading as found in the LOS case. Let $\sigma^2$ be the thermal noise power normalized by $P_\mathrm{t}$. Based on the assumptions thus far, the \ac{SINR} received by the typical user can be expressed as
\begin{align}\label{eqn:sinr}
\mbox{SINR}=\frac{\left|h_0\right|^2M_\mathrm{r}M_\mathrm{t}L(R_0)}{\sigma^2+\sum_{\ell>0:X_\ell\in\Phi}\left|h_\ell\right|^2 D_\ell L(R_\ell)}.
\end{align}
Note that the SINR in (2) is a random variable, due to the randomness in the base station locations $R_\ell$, small-scale fading $h_\ell$, and the directivity gain $D_\ell$. Using the proposed system model, we will evaluate the mmWave \ac{SINR} and rate coverage in the following section.

\section{Coverage and Rate Analysis in General Networks}\label{sec:generalnetwork}
In this section, we analyze the coverage and rate in the proposed model of a general \ac{mmWave} network. First, we provide some \ac{SINR} ordering results regarding different parameters of the antenna pattern. Then we derive expressions
for the SINR and rate coverage probability in mmWave networks with general \ac{LOS} probability function $p(R)$. To simplify subsequent analysis, we then introduce a systematic approach to approximate $p(R)$ by a moment matched equivalent step function.

\subsection{Stochastic Ordering of SINR With Different Antenna Geometries}\label{sec:geometry}
One differentiating feature of \ac{mmWave} cellular networks is the deployment of directional antenna arrays. Consequently, the performance of \ac{mmWave} networks will depend on the adaptive array pattern through the beamwidth, the directivity gain, and the back lobe gain. In this section, we establish some results on stochastic ordering of the SINRs in the systems with different antenna geometries. While we will focus on the array geometry at the transmitter, the same results, however, also apply to the receiver array geometry. The concept of stochastic ordering has been applied in analysis of wireless systems \cite{Tepedelenlioglu2011,Dhillon2013}. Mathematically, the ordering of random variables can be defined as follows \cite{Tepedelenlioglu2011,Dhillon2013}.

\begin{definition}\label{defn:order}
Let $X$ and $Y$ be two random variables. $X$ stochastically dominates $Y$, i.e., $X$ has a better distribution than $Y$, if $
\mathbb{P}(X>t)>\mathbb{P}(Y>t)
$ for all $t\in\mathbb{R}$.
\end{definition}

Next, define the \ac{FBR} at the transmitter $\xi_\mathrm{t}$ as the ratio between the main lobe directivity gain  $M_\mathrm{t}$ and the back lobe gain $m_\mathrm{t}$, i.e., $\xi_\mathrm{t}=M_\mathrm{t}/m_\mathrm{t}$. We introduce the key result on stochastic ordering of the SINR with respect to the directivity gains as follows.
\begin{proposition}[Stochastic ordering w.r.t. directivity gains]
Given a fixed beamwidth $\theta_\mathrm{t}$ and \ac{FBR} $\xi_\mathrm{t}$ at the transmitter, the mmWave network with the larger main lobe directivity gain $M_\mathrm{t}$ has a better SINR distribution. Similarly, with fixed beamwidth $\theta_\mathrm{t}$ and main lobe gain $M_\mathrm{t}$, a larger FBR $\xi_\mathrm{t}$ provides a better SINR distribution.
\end{proposition}
\proof From Definition \ref{defn:order}, we need to show that for each realization of base station locations $R_\ell$, small-scale fading $h_\ell$, and angles $\phi^\ell_{\mathrm{r}}$ and $\phi^\ell_{\mathrm{t}}$, the value of the SINR increases with $M_\mathrm{t}$ and $\xi_\mathrm{t}$. Given $R_\ell$, $h_\ell$, $\phi^\ell_{\mathrm{r}}$, and $\phi^\ell_{\mathrm{r}}$ $(\ell\in\mathbb{N})$, we can normalize both the numerator and denominator of (\ref{eqn:sinr}) by $M_t$, and then write 
$\mbox{SINR}=\frac{\left|h_0\right|^2M_\mathrm{r}L(R_0)}{\sigma^2/M_\mathrm{t}+\sum_{\ell>0:X_\ell\in\Phi}\bar{D}_\ell(\xi_\mathrm{t})\left|h_\ell\right|^2 L(R_\ell)},$
where $\bar{D}_\ell(\xi_\mathrm{t})=e_k$ with probability $b_k$, and $b_k$, $e_k$ are constants defined in Table \ref{table:constant}. Note that $\bar{D}_\ell(\xi_\mathrm{t})$ is independent of $M_\mathrm{t}$, and is a non-increasing function of $\xi_\mathrm{t}$. Hence, when $\xi_\mathrm{t}$ is fixed, larger $M_\mathrm{t}$ provides larger \ac{SINR}; when $M_\mathrm{t}$ is fixed, larger $\xi_\mathrm{t}$ provides larger \ac{SINR}. \endproof

Next, we provide the stochastic ordering result regarding beamwidth as follows.
\begin{proposition}[Stochastic ordering w.r.t. beamwidth]\label{prop:2}
Given a fixed main lobe gain $M_\mathrm{t}$ and \ac{FBR} $\xi_\mathrm{t}$ at the transmitter, a smaller beamwidth $\theta_\mathrm{t}$ provides a better \ac{SINR} distribution.
\end{proposition}

The proposition can be rigorously proved using coupling techniques. We omit the proof here and instead provide an intuitive explanation as below. Intuitively, with narrower main lobes, fewer base stations will transmit interference to the typical user via their main lobes, which gives a smaller interference power. The desired signal term in (\ref{eqn:sinr}) is independent of the beamwidth, as we ignore the channel estimation errors and potential angle spread. Hence, based on our model assumptions, smaller beamwidths provide a better SINR performance.

We note that the ordering result in Proposition \ref{prop:2} assumes that there is no angle spread in the channel. With angle spread, a narrow-beam antenna may capture only the signal energy arriving inside its main lobe, missing the energy spread outside, which causes a gain reduction in the signal power \cite{Greenstein1999}. Consequently, the results in Proposition \ref{prop:2} should be interpreted as applying to the case where beamwidths are larger than the angle spread, e.g. if the beamwidth is more than $55^\circ$ per the measurements in \cite{Rajagopal2012}. We defer more detailed treatment of angle spread to future work.

\subsection{SINR Coverage Analysis}\label{sec:analysis}\label{sec:coverageanalysis}
The \ac{SINR} coverage probability $P_\mathrm{c}(T)$ is defined as the probability that the received \ac{SINR} is larger than some threshold $T>0$, i.e.,
$
P_\mathrm{c}(T)=\mathbb{P}(\mbox{SINR}>T).
$
We present the following lemmas before introducing the main results on SINR coverage. By Assumption \ref{assump:LOS}, the outdoor base station process $\Phi$ can be divided into two independent non-homogeneous PPPs: the \ac{LOS} base station process $\Phi_\mathrm{L}$ and \ac{NLOS} process $\Phi_\mathrm{N}$. We will equivalently consider $\Phi_\mathrm{L}$ and $\Phi_\mathrm{N}$ as two independent tiers of base stations. As the user is assumed to connect to the base station with the smallest path loss, the serving base station can only be either the nearest base station in $\Phi_\mathrm{L}$ or the nearest one in $\Phi_\mathrm{N}$. The following lemma provides the distribution of the distance to the nearest base station in $\Phi_\mathrm{L}$ and $\Phi_\mathrm{N}$.

\begin{lemma}\label{lem:distance}
Given the typical user observes at least one LOS base station, the conditional probability density function of its distance to the nearest \ac{LOS} base station is
\begin{align}
f_\mathrm{L}(x)=2\pi\lambda xp(x)\mathrm{e}^{-2\pi\lambda\int_{0}^{x}rp(r)\mathrm{d}r}/B_\mathrm{L},
\end{align}
where $x>0$, $B_\mathrm{L}=1-\mathrm{e}^{-2\pi\lambda\int_{0}^{\infty}rp(r)\mathrm{d}r}$ is the probability that a user has at least one LOS base station, and $p(r)$ is the \ac{LOS} probability function defined in Section \ref{sec:system}. Similarly, given the user observes at least one NLOS base station, the conditional probability density function of the distance to the nearest \ac{NLOS} base station is
\begin{align}
f_\mathrm{N}(x)=2\pi\lambda x(1-p(x))\mathrm{e}^{-2\pi\lambda\int_{0}^{x}r(1-p(r))\mathrm{d}r}/B_\mathrm{N},
\end{align}
where $x>0$, and $B_\mathrm{N}=1-\mathrm{e}^{-2\pi\lambda\int_{0}^{\infty}r(1-p(r))\mathrm{d}r}$ is the probability that a user has at least one NLOS base station.
\end{lemma}
\proof
The proof follows  \cite[Theorem 10]{bai2013} and is omitted here.
\endproof

Next, we compute the probability that the typical user is associated with either a \ac{LOS} or a \ac{NLOS} base station.
\begin{lemma}\label{lem:prob}
The probability that the user is associated with a \ac{LOS} base station is
\begin{align}
A_\mathrm{L}=B_\mathrm{L}\int_{0}^{\infty}\mathrm{e}^{-2\pi\lambda\int_{0}^{\psi_\mathrm{L}(x)}(1-p(t))t\mathrm{d}t}f_\mathrm{L}(x)\mathrm{d}x,
\end{align}
where $\psi_\mathrm{L}(x)=\left(C_\mathrm{N}/C_\mathrm{L}\right)^{1/\alpha_\mathrm{N}}x^{\alpha_\mathrm{L}/\alpha_\mathrm{N}}$.
The probability that the user is associated with a \ac{NLOS} base station is $A_\mathrm{N}=1-A_\mathrm{L}$.
\end{lemma}
\proof See Appendix B. \endproof

Further, conditioning on that the serving base station is \ac{LOS} (or NLOS), the distance from the user to its serving base station follows the distribution given in the following lemma.
\begin{lemma}\label{lem:neardis}
Given that a user is associated with a \ac{LOS} base station, the probability density function of the distance to its serving base station is
\begin{align}
\hat{f}_\mathrm{L}(x)=\frac{B_\mathrm{L}f_\mathrm{L}(x)}{A_\mathrm{L}}\mathrm{e}^{-2\pi\lambda\int_{0}^{\psi_\mathrm{L}(x)}(1-p(t))t\mathrm{d}t},
\end{align}
when $x>0$. Given the user is served by a \ac{NLOS} base station, the probability density function of the distance to its serving base station is
\begin{align}
\hat{f}_\mathrm{N}(x)=\frac{B_\mathrm{N}f_\mathrm{N}(x)}{A_\mathrm{N}}\mathrm{e}^{-2\pi\lambda\int_{0}^{\psi_\mathrm{N}(x)}p(t)t\mathrm{d}t},
\end{align}
where $x>0$, and $\psi_\mathrm{N}(x)=\left(C_\mathrm{L}/C_\mathrm{N}\right)^{1/\alpha_\mathrm{L}}x^{\alpha_\mathrm{N}/\alpha_\mathrm{L}}$.
\end{lemma}
\proof The proof follows a similar method as that of Lemma \ref{lem:prob}, and is omitted here. \endproof

Now, based on Lemma \ref{lem:prob} and Lemma \ref{lem:neardis}, we present the main theorem on the \ac{SINR} coverage probability as follows
\begin{theorem}\label{thm:coverage}
The \ac{SINR} coverage probability $P_c(T)$ can be computed as
\begin{align}
P_\mathrm{c}(T)=A_\mathrm{L}P_{\mathrm{c},\mathrm{L}}(T)+A_\mathrm{N}P_{\mathrm{c},\mathrm{N}}(T),
\end{align}
where for $\mathrm{s}\in\{\mathrm{L},\mathrm{N}\}$, $P_{\mathrm{c},s}(T)$ is the conditional coverage probability given that the user is associated with a base station in $\Phi_\mathrm{s}$. Further, $P_{\mathrm{c},\mathrm{s}}(T)$ can be evaluated as
\begin{align}
P_{c,\mathrm{L}}(T)&\approx\sum_{n=1}^{N_\mathrm{L}}(-1)^{n+1}{N_\mathrm{L}\choose{n}}\nonumber\\\times&\int_{0}^{\infty}\mathrm{e}^{-\frac{n \eta_\mathrm{L} x^{\alpha_\mathrm{L}}T\sigma^2}{C_\mathrm{L}M_\mathrm{r}M_\mathrm{t}}-Q_n(T,x)-V_n(T,x)}\hat{f}_\mathrm{L}(x)\mathrm{d}x,\label{eqn:pc1}
\end{align}
and
\begin{align}
P_{c,\mathrm{N}}(T)&\approx\sum_{n=1}^{N_\mathrm{N}}(-1)^{n+1}{N_\mathrm{N}\choose{n}}\nonumber\\\times&\int_{0}^{\infty}\mathrm{e}^{-\frac{n \eta_\mathrm{N} x^{\alpha_\mathrm{N}}T\sigma^2}{C_\mathrm{N} M_\mathrm{r}M_\mathrm{t}}-W_n(T,x)-Z_n(T,x)}\hat{f}_\mathrm{N}(x)\mathrm{d}x.\label{eqn:pc2}
\end{align}
where
\begin{align}
Q_{n}(T,x)=2\pi\lambda\sum_{k=1}^{4} b_k \int_{x}^{\infty}F\left(N_\mathrm{L},\frac{n\eta_\mathrm{L}\bar{a}_kTx^{\alpha_\mathrm{L}}}{N_\mathrm{L}t^{\alpha_\mathrm{L} }}\right)p(t)t\mathrm{d}t,
\end{align}
\begin{align}
V_{n}(T,x)=&2\pi\lambda\sum_{k=1}^{4} b_k \int_{\psi_\mathrm{L}(x)}^{\infty}F\left(N_\mathrm{N},\frac{nC_\mathrm{N}\eta_\mathrm{L}\bar{a}_kTx^{\alpha_\mathrm{L}}}{C_\mathrm{L}N_\mathrm{N}t^{\alpha_\mathrm{N}}}\right)\nonumber\\&(1-p(t))t\mathrm{d}t,
\end{align}
\begin{align}
W_{n}(T,x)=&2\pi\lambda\sum_{k=1}^{4} b_k \int_{\psi_\mathrm{N}(x)}^{\infty}F\left(N_\mathrm{L},\frac{nC_\mathrm{L}\eta_\mathrm{N}\bar{a}_kTx^{\alpha_\mathrm{N}}}{C_\mathrm{N} N_\mathrm{L}t^{\alpha_\mathrm{L}}}\right)\nonumber\\&p(t)t\mathrm{d}t,
\end{align}
\begin{align}
Z_{n}(T,x)=&2\pi\lambda\sum_{k=1}^{4} b_k \int_{x}^{\infty}F\left(N_\mathrm{N},\frac{n\eta_\mathrm{N}\bar{a}_kTx^{\alpha_\mathrm{N}}}{N_\mathrm{N}t^{\alpha_\mathrm{N}}}\right)\nonumber\\&(1-p(t))t\mathrm{d}t,
\end{align}
and $F(N,x)=1-1/(1+x)^N$. For $s\in\{\mathrm{L},\mathrm{N}\}$, $\eta_s=N_s(N_{s}!)^{-\frac{1}{N_s}}$, $N_{s}$ are the parameters of the Nakagami small-scale fading; for $k\in\{1,2,3,4\}$, $\bar{a}_k=\frac{a_k}{M_\mathrm{t}M_\mathrm{r}}$, $a_k$ and $b_k$ are constants defined in Table \ref{table:constant}.
\end{theorem}
\proof
See Appendix C.
\endproof
  Though as an approximation of the SINR coverage probability, we find that the expressions in Theorem \ref{thm:coverage} compare favorably with the simulations in Section \ref{sec:simu}. In addition,  the expressions in Theorem \ref{thm:coverage} compute much more efficiently than prior results in \cite{bai2013b}, which required a numerical inverse of a Fourier transform. Last, the LOS probability function $p(t)$ may itself have a very complicated form, e.g. the empirical function for small cell simulations in \cite{3GPPTR36.8142010}, which will make the numerical evaluation difficult. Hence, we propose simplifying the system model by using a step function to approximate $p(t)$ in Section $\ref{sec:LOS}$. Before that, we introduce our rate analysis results in the following section.

\subsection{Rate Analysis}
In this section, we analyze the distribution of the achievable rate $\Gamma$ in \ac{mmWave} networks. We use the following definition for the achievable rate
\begin{align}
\Gamma=W\log_2(1+\min\{\mbox{SINR},T_{\max}\}),
\end{align}
where $W$ is the bandwidth assigned to the typical user, and $T_{\max}$ is a SINR threshold determined by the order of the constellation and the limiting distortions from the RF circuit. The use of a distortion threshold $T_{\max}$ is needed because of the potential for very high SINRs in \ac{mmWave} that may not be exploited due to other limiting factors like linearity in the radio frequency front-end.

The  average achievable rate $\mathbb E[\Gamma]$ can be computed using the following Lemma from the \ac{SINR} coverage probability $P_\mathrm{c}(T)$.
\begin{lemma}\label{lem:avgrate}
Given the \ac{SINR} coverage probability $P_\mathrm{c}(T)$, the average achievable rate in the network is
$
\mathbb E\left[\Gamma\right]=\frac{W}{\ln 2}\int_{0}^{T_{\max}}\frac{P_\mathrm{c}(T)}{1+T}\mathrm{d}T.
$
\end{lemma}
\proof See \cite[Theorem 3]{Andrews2011a} and \cite [Section V]{akoum2012}. \endproof 

Lemma \ref{lem:avgrate} provides a first order characterization of the rate distribution. We can also derive the exact rate distribution using the  {\it rate coverage probability} $P_\mathrm{R}(\gamma)$, which is the probability that the achievable rate of the typical user is larger than some threshold $\gamma$:
$
P_\mathrm{R}(\gamma)=\mathbb P[\Gamma>\gamma].
$
The rate coverage probability $P_\mathrm{R}(\gamma)$ can be evaluated through a change of variables as in the following lemma.
\begin{lemma}\label{lemma:rate}
Given the \ac{SINR} coverage probability $P_\mathrm{c}(T)$, for $\gamma<W\log_\mathrm{N}(1+T_{\max})$, the rate coverage probability can be computed as
$
P_\mathrm{R}(\gamma)=P_\mathrm{c}(2^{\gamma/W}-1).
$
\end{lemma}
\proof The proof is similar to that of \cite[Theorem 1]{singh2013}. For $\gamma<W\log_\mathrm{N}(1+T_{\max})$, it directly follows that
$P_\mathrm{R}(\gamma)=\mathbb P\left[\mbox{SINR}>2^{\gamma/W}-1\right]=P_\mathrm{c}(2^{\gamma/W}-1)$.\endproof 
Lemma \ref{lemma:rate} will allow comparisons to be made between mmWave and conventional systems that use different bandwidths, as presented in Section \ref{sec:simu}.

\subsection{Simplification of \ac{LOS} Probability Function}\label{sec:LOS}
\begin{figure}[!ht]
\centering
\subfigure[center][{Irregular shape of an acutal LOS region.}]{
\includegraphics[width=0.7\columnwidth]{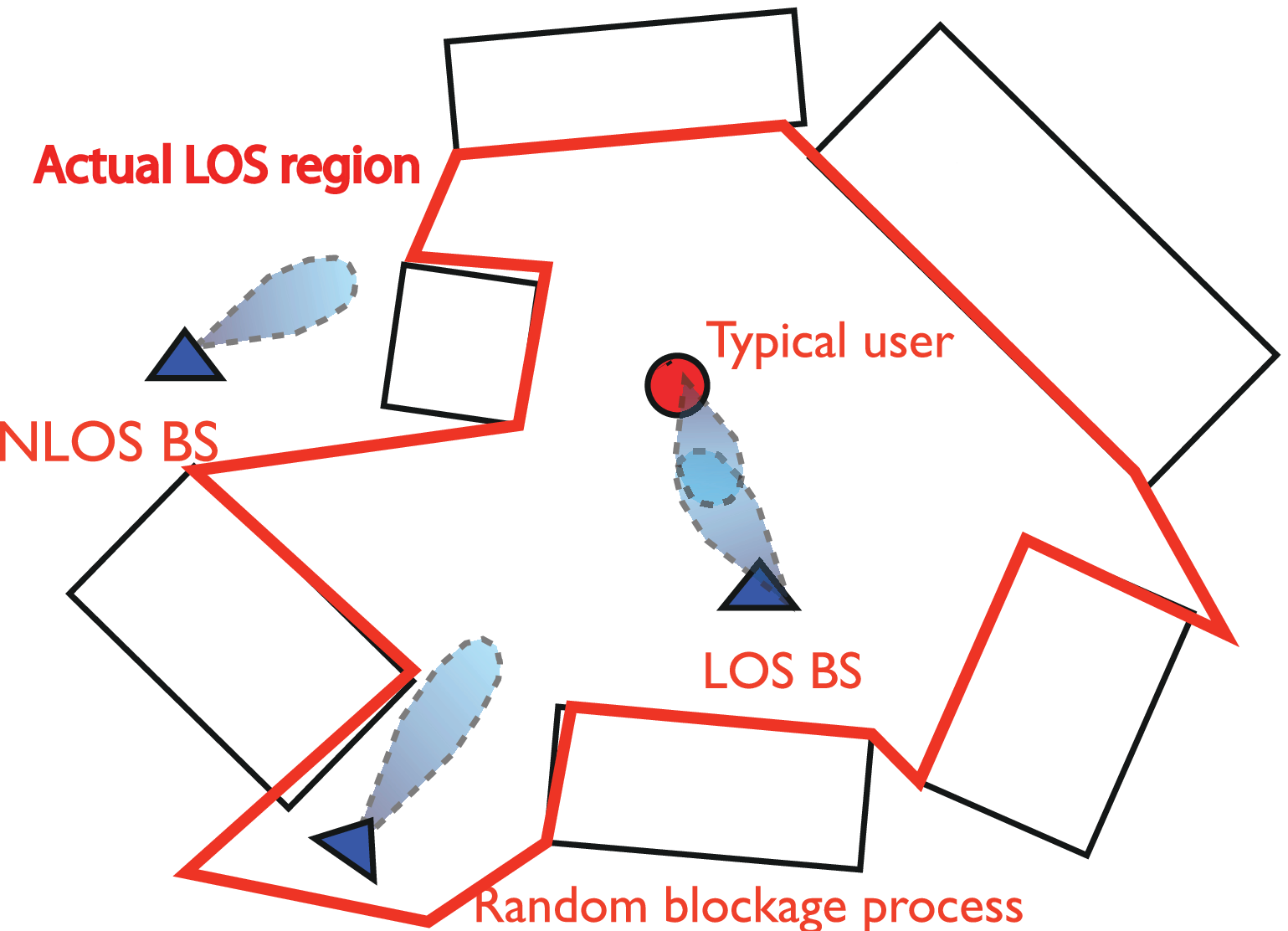}}
\subfigure[center][{Approximation using the equivalent \ac{LOS} ball.}]{
\includegraphics[width=0.7\columnwidth]{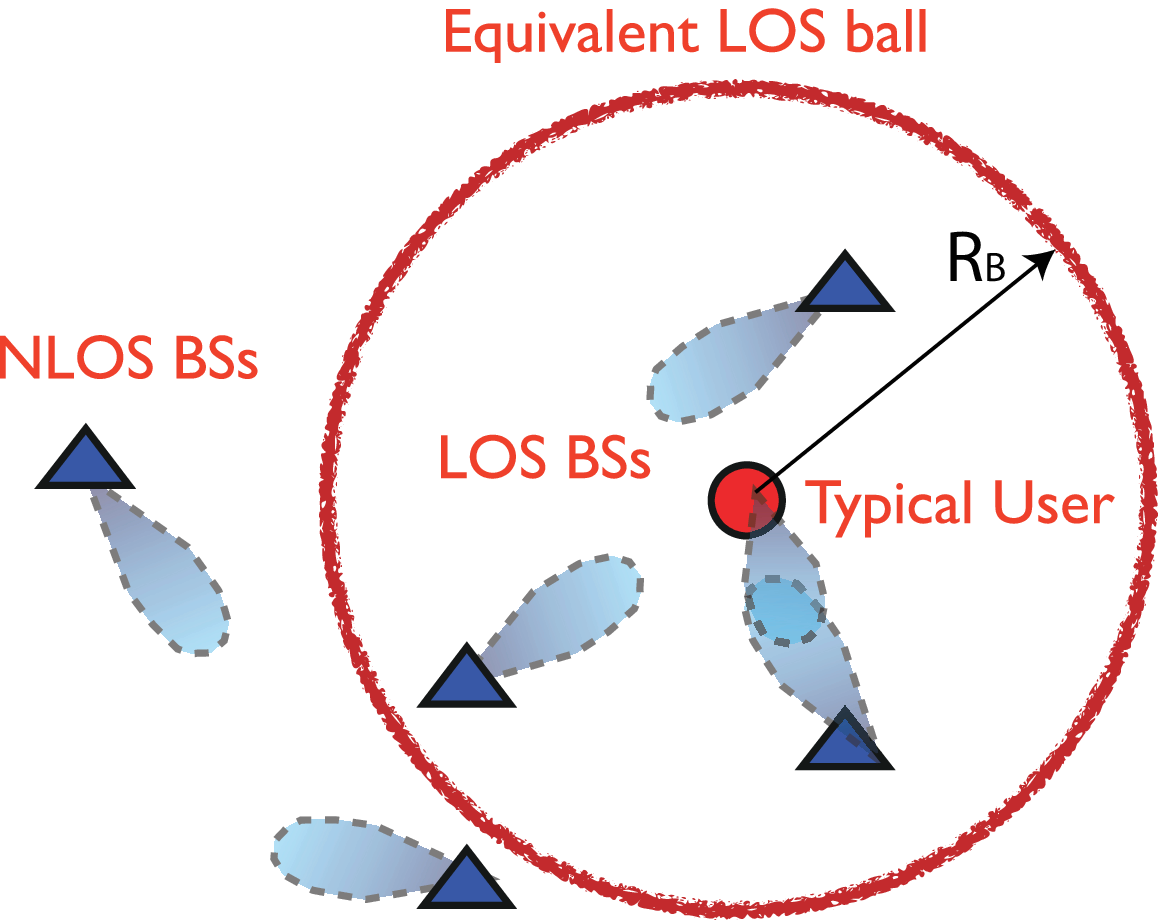}}
\caption{ Simplification of the random \ac{LOS} region as a fixed equivalent \ac{LOS} ball. In (a), we illustrate one realization of randomly located buildings corresponding to a general \ac{LOS} function $p(x)$. The \ac{LOS} region observed by the typical user has an irregular shape. In (b), we approximate $p(x)$ by a step function. Equivalently, the \ac{LOS} region is also approximated by a fixed ball. Only base stations inside the ball are considered LOS to the user.}\label{fig:LOSball}
\end{figure}
The expressions in Theorem \ref{thm:coverage} generally require numerical evaluation of multiple integrals, and may become difficult to analyze. In this section, we propose to simplify the analysis by approximating a general \ac{LOS} probability function $p(t)$ by a step function. We denote the step function as $S_{R_\mathrm{B}}(x)$, where $S_{R_\mathrm{B}}(x)=1$ when $0<x<R_\mathrm{B}$, and $S_{R_\mathrm{B}}(x)=0$ otherwise. Essentially, the LOS probability of the link is taken to be one within a certain fixed radius $R_\mathrm{B}$ and zero outside the radius. An interpretation of the simplification is that the irregular geometry of the LOS region in Fig. \ref{fig:LOSball} (a) is replaced with its equivalent \ac{LOS} ball in Fig. \ref{fig:LOSball} (b). Such simplification not only provides efficient expressions to compute \ac{SINR}, but enables simpler analysis of the network performance when the network is dense.

We will propose two criterions to determine the $R_\mathrm{B}$ given \ac{LOS} probability function $p(t)$. Before that, we first review some useful facts.
\begin{theorem}\label{lem:NL}
Given the \ac{LOS} probability function $p(x)$, the average number of \ac{LOS} base stations that a typical user  observes is $\rho=2\pi\lambda\int_{0}^{\infty}p(t)t\mathrm{d}t$.
\end{theorem}
\proof
The average number of \ac{LOS} base stations can be computed as
\begin{align*}
\rho=\mathbb E\left[\sum_{X_\ell\in\Phi}\mathbb{I}(X_\ell\in\Phi_\mathrm{L})\right]
\stackrel{(a)}=2\pi\lambda\int_{0}^{\infty}p(t)t\mathrm{d}t,
\end{align*}
where (a) follows directly from Campbell's formula of PPP \cite{BacBla:Stochastic-Geometry-and-Wireless:09}.
\endproof
A direct corollary of Theorem \ref{lem:NL} follows as below.
\begin{corollary}\label{cor:NL1}
When $p(x)=S_{R}(x)$, the average number of LOS base stations is $\rho=\pi\lambda R^2$.
\end{corollary}

Note that Theorem \ref{lem:NL} also indicates that a typical user will observe a finite number of \ac{LOS} base stations almost surely when $\int_{0}^{\infty}p(t)t\mathrm{d}t<\infty$. Hence, if $p(x)$ satisfies $\int_{0}^{\infty}p(t)t\mathrm{d}t<\infty$,
 the parameter $R_\mathrm{B}$ in $S_{R_\mathrm{B}}(x)$ can be determined by matching the average number of \ac{LOS} base stations a user may observe.
\begin{criterion}[Mean \ac{LOS} BS Number]\label{crit:NL}
When $\int_{0}^{\infty}p(t)t\mathrm{d}t<\infty$, the parameter $R_\mathrm{B}$ of the equivalent step function $S_{R_\mathrm{B}}(x)$ is determined to match the first moment of $\rho$. By Theorem \ref{lem:NL}, it follows that $R_\mathrm{B}=\left(2\int_{0}^{\infty}p(t)t\mathrm{d}t\right)^{0.5}$.
\end{criterion}

In the case where $\int_{0}^{\infty}p(t)t\mathrm{d}t<\infty$ is not satisfied, another criterion to determine $R_\mathrm{B}$ is needed. Note that even if the first moment is infinite, the probability that the user is associated with a \ac{LOS} base station exists and is naturally finite for all $p(t)$. Hence, we propose the second criterion regarding the \ac{LOS} association probability as follows.
\begin{criterion}[LOS Association Probability]\label{crit:prob}
Given a \ac{LOS} probability function $p(t)$, the parameter $R_\mathrm{B}$ of its equivalent step function $S_{R_\mathrm{B}}(x)$ is determined such that the \ac{LOS} association probability $A_\mathrm{L}$ is unchanged after approximation.
\end{criterion}
From Lemma \ref{lem:prob}, the \ac{LOS} association probability for a step function $S_{R_\mathrm{B}}(x)$ equals $1-\mathrm{e}^{-\lambda\pi R_\mathrm{B}^2}$. Hence, by Criterion \ref{crit:prob}, $R_\mathrm{B}$ can be determined as $R_\mathrm{B}=\left(\frac{-\ln (1-A_\mathrm{L})}{\lambda \pi}\right)^{0.5}$.

Last, we explain the physical meaning of the step function approximation as follows. As shown in Fig. \ref{fig:LOSball}(a), with a general \ac{LOS} probability function $p(x)$, the buildings are randomly located, and thus the actual \ac{LOS} region observed by the typical user may have an unusual shape. Although it is possible to incorporate such randomness of the size and shape by integrating over $p(t)$, the expressions with multiple integrals can make the analysis and numerical evaluation difficult \cite{bai2013b}. In Fig. \ref{fig:LOSball}(b), by approximating the \ac{LOS} probability function as a step function $S_{R_\mathrm{B}}(x)$, we equivalently approximate the \ac{LOS} region by a fixed ball $\mathcal{B}(0,R_\mathrm{B})$, which we define as the {\it equivalent \ac{LOS} ball}. As will be shown in Section \ref{sec:dense}, approximating $p(x)$ as a step function enables fast numerical computation, simplifies the analysis, and provides design insights for dense network. Besides, we will show in simulations in Section \ref{sec:simu} that the error due to such approximation is generally small in dense mmWave networks, which also motivates us to use this first-order approximation of the LOS probability function to simplify the dense network analysis in the following section.

\section{Analysis of Dense \ac{mmWave} Networks}\label{sec:dense}

In this section we specialize our results to dense networks. This approach is motivated by subsequent numerical results in Section \ref{sec:simu} that show \ac{mmWave} deployments will be dense if they are expected to achieve significant coverage. We  derive simplified expressions for the \ac{SINR} and provide further insights into system performance in this important asymptotic regime.

\subsection{Dense Network Model}
In this section, we build the dense network model by modifying the system model in Section \ref{sec:system} with a few additional assumptions. We say that a  \ac{mmWave} cellular network is {\it dense} if the average number of LOS base stations observed by the typical user $\rho$ is larger than $K$, or if its LOS association probability $A_\mathrm{L}$ is larger than $1-\epsilon$, where $K$ and $\epsilon$ are pre-defined positive thresholds. In this paper, for illustration purpose, we will let $K=1$ and $\epsilon=5\%$. Further, we say that a network is {\it ultra-dense} when $\rho>10$. Note that $\rho$ also equals the relative base station density normalized by the average LOS area, in this special case, as we will explain below.

Now we make some additional assumptions that will allow us to further simplify the network model.
\begin{assumption}[LOS equivalent ball]\label{assump:LOSball}
The \ac{LOS} region of the typical user is approximated by its equivalent \ac{LOS} ball $\mathcal{B}(0,R_\mathrm{B})$ as defined in Section \ref{sec:LOS}.
\end{assumption}

By Assumption \ref{assump:LOSball}, the \ac{LOS} probability function $p(t)$ is approximated by its equivalent step function $S_{R_\mathrm{B}}(x)$, and the \ac{LOS} base station process $\Phi_\mathrm{L}$ is made up of the outdoor base stations that are located inside the \ac{LOS} ball $\mathcal{B}(0,R_\mathrm{B})$. Noting that the outdoor base station process $\Phi$ is a homogeneous PPP with density $\lambda$, the average number of LOS base stations is $\rho=\lambda\pi R_\mathrm{B}^2$, which is the outdoor base station density times the area of the LOS region. For  ease of illustration, we call $\rho$ the {\it the relative density} of a mmWave network. The relative density $\rho$ is equivalently: (i) the average number of \ac{LOS} base stations that a user will observe, (ii) the ratio of the average LOS area $\pi R_\mathrm{B}^2$ to the size of a typical cell $1/\lambda$ \cite{BacBla:Stochastic-Geometry-and-Wireless:09}, and (iii) the normalized base station density by the size of the LOS ball. We will show in the next section that the SINR coverage
in dense networks is largely determined by the relative density $\rho$.

\begin{assumption}[No NLOS and noise] Both NLOS base stations and thermal noise are ignored in the analysis since in the dense regime, the performance is limited by other LOS interferers.
\end{assumption}

We show later in the simulations that ignoring NLOS base stations and the thermal noise introduces a negligible error in the performance evaluation.
\begin{assumption}[No Small-scale fading]
Small-scale fading is ignored in the dense network analysis, as the signal power from a nearby mmWave LOS transmitter is found to be almost deterministic in measurements \cite{Rappaport2013a}.
\end{assumption}

Based on the dense network model, the signal-to-interference ratio (SIR) can be expressed as
\begin{align}\label{eqn:SIR}
\mbox{SIR}=\frac{M_\mathrm{t}M_\mathrm{r} R_0^{-\alpha_\mathrm{L}}}{\sum_{\ell:X_\ell\in\Phi\cap\mathcal{B}(0,R_\mathrm{B})}D_\ell R_\ell^{-\alpha_\mathrm{L}}}.
\end{align}
Now we compute the SIR distribution in the dense network model.

\subsection{Coverage Analysis in Dense Networks}
Now we present  an approximation of the SINR distribution in a mmWave dense network. Our main result is summarized in the following theorem.

\begin{theorem}\label{thm:dense}The \ac{SINR} coverage probability in a dense network can be approximated as
{\begin{align}
P_\mathrm{c}(T)&\approx\rho\mathrm{e}^{-\rho}\sum_{\ell=1}^{N}(-1)^{\ell+1}{N\choose{\ell}}\int_{0}^{1}\prod_{k=1}^{4}\exp\left(-\frac{2}{\alpha_\mathrm{L}}b_k\rho t\right.\nonumber\\&\left.\times \left(\ell\eta T\bar{a}_k\right)^{\frac{2}{\alpha_\mathrm{L}}}\Gamma\left(-\frac{2}{\alpha_\mathrm{L}};\ell\eta T\bar{a}_k,\ell\eta T\bar{a}_kt^{\frac{\alpha_\mathrm{L}}{2}}\right)\right)\mathrm{d}t,\label{eqn:thmdense1}
\end{align}}
where $\Gamma(s;a,b)=\int_{a}^{b}x^{s-1}\mathrm{e}^{-x}\mathrm{d}x$ is the incomplete gamma function, $\bar{a}_k=a_k/(M_\mathrm{t}M_\mathrm{r})$, $a_k$ and $b_k$ are defined in Table \ref{table:constant}, $\eta=N(N!)^{\frac{1}{N}}$, and $N$ is the number of terms used in the approximation.
\end{theorem}
\proof See Appendix D.\endproof
When $\alpha_\mathrm{L}=2$, the expression in Theorem \ref{thm:dense} can be further simplified as follows.
\begin{corollary}
When $\alpha_\mathrm{L}=2$, the \ac{SINR} coverage probability approximately equals 
\begin{align}
&P_\mathrm{c}(T)\approx\rho\mathrm{e}^{-\rho}\sum_{\ell=1}^{N}(-1)^{\ell+1}{N\choose{\ell}}\int_{0}^{1}\prod_{k=1}^{4}\exp\left(\rho b_k\times\right.\nonumber\\&\left.\left(\mathrm{e}^{-\ell\eta T\bar{a}_k t}-t\mathrm{e}^{-\ell\eta T\bar{a}_k}\right)\right)\left(\frac{1-\mathrm{e}^{-\ell\mu\eta T \bar{a}_k  t}}{1-\mathrm{e}^{-\ell\mu\eta T \bar{a}_k}}\right)^{\ell\eta T b_k \bar{a}_k\rho t}\mathrm{d}t,\label{eqn:thmdense2}
\end{align}
where $\mu=\mathrm{e}^{0.577}$.
\end{corollary}

The results in Theorem \ref{thm:dense} generally provide a close approximation of the SINR distribution when enough terms are used, e.g. when $N\ge5$, as will be shown in Section \ref{sec:simu2}. More importantly, we note that the expressions in Theorem \ref{thm:asymp} are very efficient to compute, as most numerical tools support fast evaluation of the gamma function in (\ref{eqn:thmdense1}), and (\ref{eqn:thmdense2}) only requires a simple integral over a finite interval. Besides, given the path loss exponent $\alpha_\mathrm{L}$ and the antenna geometry $a_k$, $b_k$, Theorem \ref{thm:dense} shows that the approximated SINR is only a function of the relative density $\rho$, which indicates the SIR distribution in a dense network is mostly determined on the average number of LOS base station to a user.

\subsection{Asymptotic Analysis in Ultra-Dense Networks}\label{sec:asymp} 
To obtain further insights into coverage in dense networks, we provide results on the asymptotic SIR distribution when the relative density $\rho$ becomes large. We use this distribution to answer the following questions: (i) What is the asymptotic SIR distribution when the network becomes extremely dense? (ii) Does increasing base station density always improve SIR in a mmWave network?


First, we present the main asymptotic results as follows.
\begin{theorem}\label{thm:asymp}
In a dense network, when the LOS path loss exponent $\alpha_\mathrm{L}\le2$, the SIR converges to zero in probability, as $\rho\to\infty$. When $\alpha_\mathrm{L}>2$, the SIR converges to a nonzero random variable $\mbox{SIR}_0$ in distribution, as $\rho\to\infty$; Based on \cite[Proposition 10]{Blaszczyszyn2013}, a lower bound of the coverage probability for the asymptotic $\mbox{SIR}_0$ is that for $T>1$, $$\mathbb{P}(\mbox{SIR}_0>T)\ge\frac{\alpha_\mathrm{L} T^{-2/\alpha_\mathrm{L}}}{2\pi\sin(2\pi/\alpha_\mathrm{L})}.$$
\end{theorem}
\proof  See Appendix E.\endproof
Note that Theorem \ref{thm:asymp} indicates that increasing base station density above some threshold will hurt the system performance, and that the SINR optimal base station density  is finite.

Now we provide an intuitive explanation of the asymptotic results as follows. When increasing the base station density, the distances between the user and base stations become smaller, and the user becomes more likely to be associated with a \ac{LOS} base station. When the density is very high, however, a user sees several \ac{LOS} base stations and thus experiences significant interference.

We note that the asymptotic trends in Theorem \ref{thm:asymp} are valid when base stations are all assumed to be active in the network. A simple way to avoid ``over-densification'' is to simply turn off a fraction of the base stations.  This is a simple kind of interference management; study of more advanced interference management concepts is an interesting topic for future work.

\section{Numerical Simulations}
In this section, we first present some numerical results based on our analyses in Section \ref{sec:generalnetwork} and Section \ref{sec:dense}. We conclude with some simulations using real building distributions to validate our proposed mmWave network model. 
\subsection{General Network Simulations}\label{sec:simu}
In this section, we provide numerical simulations to validate our analytical results in Section \ref{sec:generalnetwork}, and further discuss their implications on system design. We assume the \ac{mmWave} network is operated at 28 GHz, and the bandwidth assigned to each user is $W=100$ MHz. The \ac{LOS} and \ac{NLOS} path loss exponents are $\alpha_\mathrm{L}=2$ and $\alpha_\mathrm{N}=4$. The parameters of the Nakagami fading are $N_\mathrm{L}=3$ and $N_\mathrm{N}=2$.  We assume the LOS probability function is $p(x)=\mathrm{e}^{-\beta x}$, where $1/\beta=141.4$ meters. For the ease of illustration, we define the notion of {\it the average cell radius} of a network as follows. Note that if the base station density is $\lambda$, the average cell size in the network is $1/\lambda$ \cite{BacBla:Stochastic-Geometry-and-Wireless:09}. Therefore, the average cell radius $r_\mathrm{c}$ in a network is defined as the radius of a ball that has the size of an average cell, i.e., $r_\mathrm{c}=\sqrt{1/\pi\lambda}$. The average cell radius not only directly relates to the inter-site distance that is used by industry in base station planning, but also equivalently characterizes the base station density in a network; as a large average cell size indicates a low base station density in the network.
\begin{figure} [!ht]
\centerline{
\includegraphics[width=0.9\columnwidth]{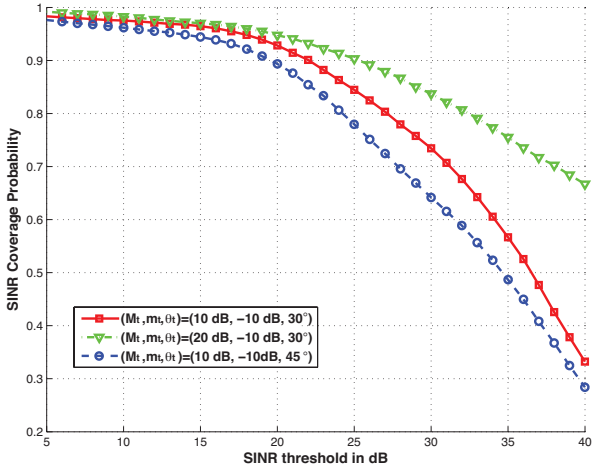}
   }
\caption{ \ac{SINR} coverage probability with different antenna geometry. The average cell radius is $r_\mathrm c=100$ meters. The receiver beam pattern is fixed as $G_{10\mbox{dB},-10\mbox{dB},90^\circ}$.}
\label{fig:geometry}
\end{figure}
\begin{figure} [!ht]
\centerline{
\includegraphics[width=0.9\columnwidth]{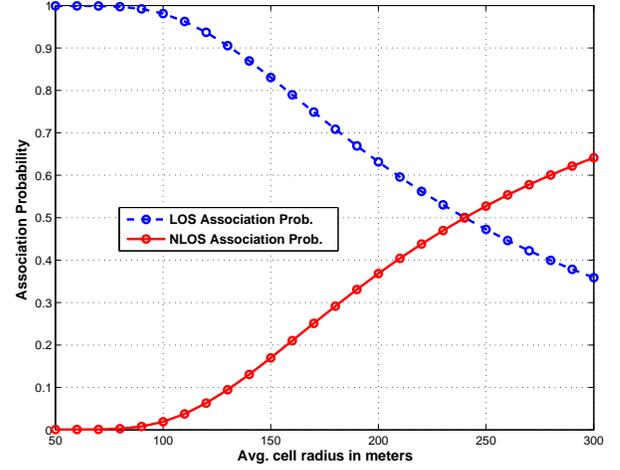}
   }
\caption{ LOS association probability with different average cell radii. The lines are drawn from Monte Carlos simulations, and the marks are drawn based on Lemma \ref{lem:prob}.}
\label{fig:LOSprob}
\end{figure}

First, we compare the \ac{SINR} coverage probabilities with different transmit antenna parameters in Fig. \ref{fig:geometry} using Monte Carlos simulations. As shown in Fig. \ref{fig:geometry}, when the side lobe gain $m_\mathrm{t}$ is fixed, better SINR performance is achieved by increasing main lobe gain $M_\mathrm{t}$ and by decreasing the main lobe beamwidth $\theta_\mathrm{t}$, as indicated by the analysis in Section \ref{sec:geometry}.

\begin{figure}[!ht]
\centering
\subfigure[center][{Analytical bounds using Theorem \ref{thm:coverage}.}]{
\includegraphics[width=0.9\columnwidth]{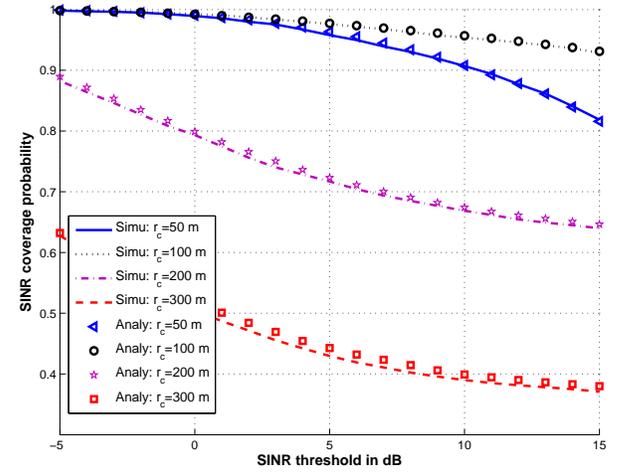}}
\subfigure[center][{Comparison between SINR and SIR.}]{
\includegraphics[width=0.9\columnwidth]{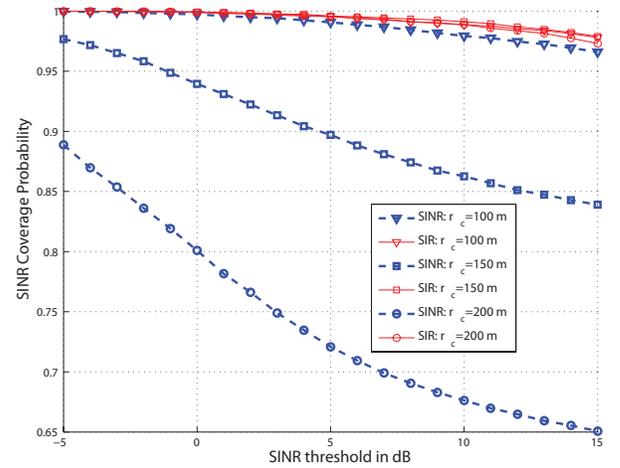}}
\caption{ SINR coverage probability with different average cell sizes. The transmit antenna pattern is assumed to be $G_{100\mbox{dB},0\mbox{dB},30^\circ}$. In (a), analytical results from Theorem \ref{thm:coverage} are shown to provide a tight approximation. In (b), it shows that SIR converges to SINR when the base station density becomes high, which implies that mmWave networks can be interference-limited.}\label{fig:density}
\end{figure}

\begin{figure} [!ht]
\centerline{
\includegraphics[width=0.9\columnwidth]{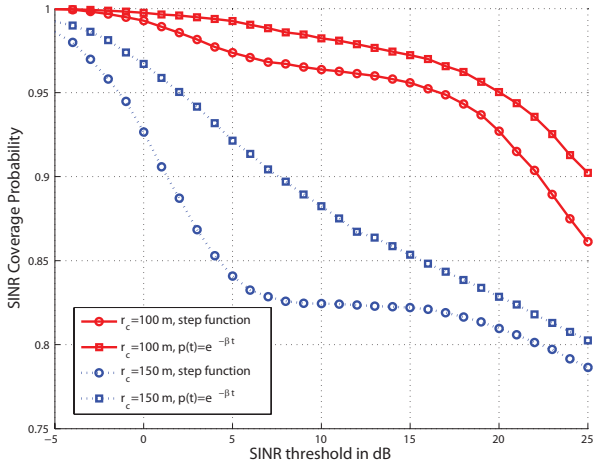}
   }
\caption{ Comparison of the \ac{SINR} coverage between using $p(x)$ and its equivalent step function $S_{R_\mathrm{B}}(x)$. The transmit antenna pattern is assumed to be $G_{20\mbox{dB},-10\mbox{dB},30^\circ}$. It shows that the step function tends to provide a more pessimistic SINR coverage probability, but the gap becomes smaller as the network becomes more dense.}
\label{fig:validate}
\end{figure}
Next, we compare the LOS association probabilities $A_\mathrm{L}$ with different average cell radii in Fig. \ref{fig:LOSprob}. The results show that the probability that a user is associated with a LOS base station increases as the cell radius decreases. The results in Fig. \ref{fig:LOSprob} also indicate that the received signal power will be mostly determined by the distribution of LOS base stations in a sufficiently dense network, e.g. when the average cell size is smaller than 100 meters in the simulation. 

We also compare the \ac{SINR} coverage probability with different cell radii in Fig. \ref{fig:density}. The numerical results in Fig. \ref{fig:density} (a) show that our analytical results in Theorem \ref{thm:coverage} match the simulations well with negligible errors. Unlike in a interference-limited conventional cellular network, where SINR is almost invariant with the base station density \cite{Andrews2011a}, the \ac{mmWave} SINR coverage probability is also shown to be sensitive to the base station density in Fig. \ref{fig:density}. The results in Fig. \ref{fig:density} (a) also shows that \ac{mmWave} networks generally require a small cell radius (equivalently a high base station density) to achieve acceptable \ac{SINR} coverage. Moreover, the results in Fig. \ref{fig:density} (b) show that when decreasing average cell radius (i.e., increasing base station density), mmWave networks will transit from power-limited regime into interference-limited regime; as the SIR curves will converge to the SINR curve when densifying the network.

Specifically, comparing the curves for $r_c=200$ meters and $r_c=300$ meters in \figref{fig:density} (a), we find that increasing base station density generally improve the SINR in a sparse network; as increasing base station density will increase the LOS association probability and avoid the presence of coverage holes, i.e. the cases that a user observes no LOS base stations. A comparison of the curves for $r_c=100$ meters and $r_c=50$ meters, however, also indicates that increasing base station density need not improve SINR, especially when the network is already sufficiently dense. Intuitively, increasing base station density also increases the likelihood to be interfered by strong LOS interferers. In a sufficiently dense network, increasing base station will harm the SINR by adding more strong interferers.


\begin{figure} [!ht]
\centerline{
\includegraphics[width=0.9\columnwidth]{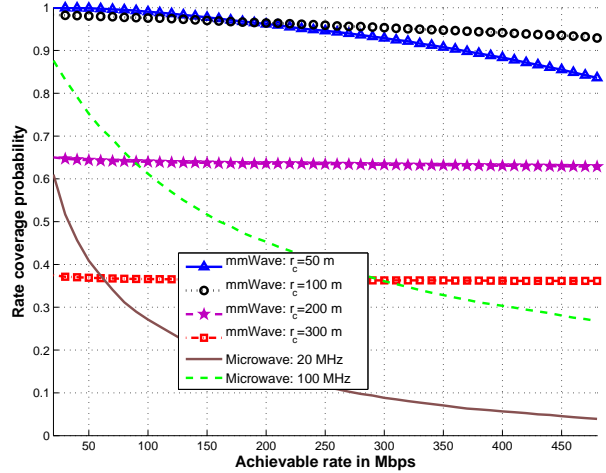}
   }
\caption{ Rate coverage comparison between mmWave and conventional cellular networks. The mmWave transmit antenna pattern is assumed to be $G_{10\mbox{dB},-10\mbox{dB},30^\circ}$. We assume the conventional system is operated at 2 GHz with a cell radius of 500 m, and the transmit power of the conventional base station is $46$ dBm.}\label{fig:rate} 
\end{figure}

Now we apply Theorem 3 to compare the SINR coverage with different LOS probability functions $p(x)$. We approximate the negative exponential function $p(x)=\mathrm{e}^{-\beta x}$ by its equivalent step function $S_{R_\mathrm{B}}(x)$. Applying either of the criteria in Section \ref{sec:LOS}, the radius of the equivalent LOS ball $R_\mathrm{B}$ equals 200 meters. As shown in Fig. \ref{fig:validate}, the step function approximation generally provides a lower bound of the actual SINR distribution, and the errors due to the approximation become smaller when the base station density increases.
The approximation of step function also enables faster evaluations of the coverage probability, as it simplifies expressions for the numerical integrals.

We provide rate results in Fig. \ref{fig:rate}, where the lines are drawn from Monte Carlos simulations, and the marks are drawn based on Lemma \ref{lemma:rate}. In the rate simulation, we assume that 64 QAM is the highest constellation supported in the networks, and thus the maximum spectrum efficiency per data stream is 6 bps/Hz.
In Fig. \ref{fig:rate}, we compare the rate coverage probability between the mmWave network and a conventional network operated at 2 GHz. The mmWave bandwidth is 100 MHz (which conceivably could be much larger, e.g. 500 MHz \cite{pi2011,akdeniz2013}), while we assume the conventional system has a basic bandwidth of 20 MHz, which can be potentially extended to 100 MHz by enabling carrier aggregation \cite{ghosh2010}. Rayleigh fading is assumed in the UHF network simulations. We further assume that conventional base stations have perfect channel state information, and apply spatial multiplexing (4$\times$4 single user MIMO with zero-forcing precoder) to transmit multiple data streams. More comparison results with other techniques can be found in \cite{bai2014a}. Results in Fig. \ref{fig:rate} shows that, due to the favorable SINR distribution and larger available bandwidth at mmWave frequencies, the mmWave system with a sufficiently small average cell size outperforms the conventional system in terms of providing high data rate coverage.
\subsection{Dense Network Simulations}\label{sec:simu2}
\begin{figure}[!ht]
\centerline{
\includegraphics[width=0.9\columnwidth]{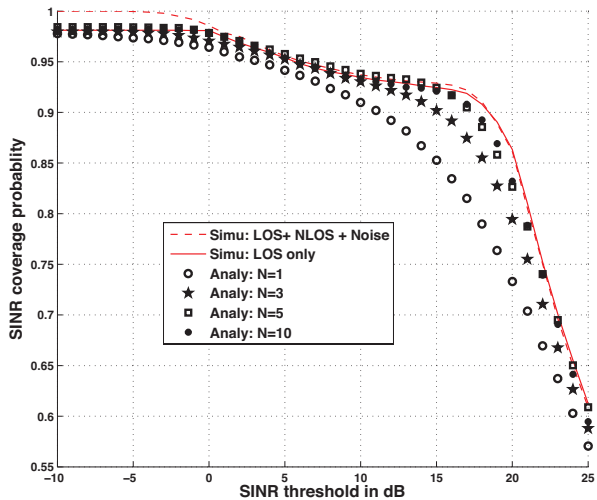}
   }
\caption{ Coverage probability in a dense mmWave network. The mmWave transmit antenna pattern is assumed to be $G_{10\mbox{dB},-10\mbox{dB},30^\circ}$. We assume $R_\mathrm{B}=200$ m, and the relative base station density $\rho=4$. $N$ is the number of terms we used to approximate the coverage probability in Theorem \ref{thm:dense}.}\label{fig:densevalidate}
\end{figure}

Now we show the simulation results based on the dense network analysis in Section \ref{sec:dense}. First, we illustrate the results in Theorem \ref{thm:dense} with the simulations in Fig. \ref{fig:densevalidate}. In the simulations, we include the \ac{NLOS} base stations and thermal noise, which were ignored in the theoretical derivation. The expression derived in Theorem \ref{thm:dense} generally provides a lower bound of the coverage probability. The approximation becomes more accurate when more terms are used in the approximation, especially when $N\ge5$. We find that the error due to ignoring NLOS base stations and thermal noise is minor in terms of the SINR coverage probability, primarily impacting low SINRs.

\begin{figure}[!ht]
\centering
\subfigure[center][{Analytical results using Theorem \ref{thm:dense}.}]{
\includegraphics[width=0.9\columnwidth]{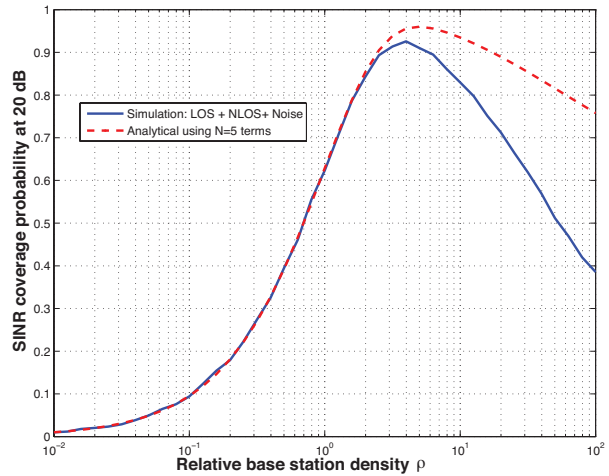}}
\subfigure[center][{{Optimal density with different path loss exponents.}}]{
\includegraphics[width=0.9\columnwidth]{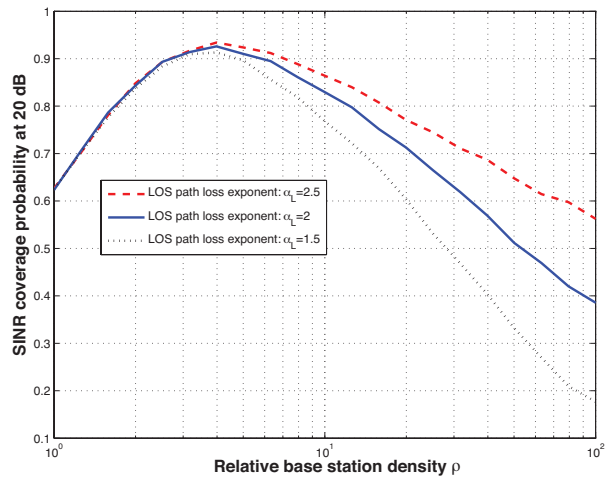}}
\caption{ SINR coverage probability with different relative base station density when the target SINR=20 dB. In the simulations, we include the \ac{NLOS} base stations outside the \ac{LOS} region and the thermal noise. We also fix the radius of the LOS ball as $R_\mathrm{B}=$200 meters, and change the base station density $\lambda$ at each step according to the value of the relative base station density $\rho$. In (a), it shows that ignoring NLOS base stations and the noise power causes minor errors in terms of the optimal cell radius. In (b), we search for the optimal relative density with different LOS path loss exponents. It shows that the optimal cell radius is generally insensitive to the path loss exponent.}\label{fig:optimal}
\end{figure}

Next, we compare the \ac{SINR} coverage probability with different relative base station density when $T=20$ dB. Recall that $\rho=\lambda\pi R_\mathrm{B}^2$ is the base station density normalized by the size of the LOS region. In \ref{fig:optimal}(a), the path loss exponent is assumed to be $\alpha=2$. We compute the coverage probability from $\rho=-20$ dB meters to $\rho=20$ dB with a step of 1 dB. The analytical expressions in Theorem \ref{thm:dense} are much more efficient than simulations: the plot takes seconds to finish using the analytical expression, while it approximately takes an hour to simulate 10,000 realizations at each step. As shown in \figref{fig:optimal} (a), although there is some gap between the simulation and the analytical results in the ultra-dense network regime, both curves achieve their maxima at approximately $\rho = 5$, i.e., when the average cell radius $r_c$ is approximately 1/2 of the LOS range $R_\mathrm{B}$. Moreover, when the base station density grows very large, the coverage probability begins to decrease, which matches the asymptotic results in Theorem \ref{thm:asymp}. The results also indicate that networks in the environments with dense blockages, e.g. the downtown areas of large cities where the LOS range $R_\mathrm{B}$ is small, will benefit from network densification; as they are mostly operated in the region where the relative density is (much) smaller than the optimal value $\rho\approx5$, and thus increasing $\rho$ by densifying networks will improve SINR coverage.

We also simulate with other LOS path loss exponents in Fig. \ref{fig:optimal} (b). The results show that the optimal base station density is generally insensitive to the change of the path loss exponent. When the LOS path loss exponent increases from 1.5 to 2.5, the optimal cell size is almost the same. The results also illustrate that the networks with larger path loss exponent $\alpha_\mathrm{L}$ have better SINR coverage in the ultra-dense regime when $\rho>10$. Intuitively, signals attenuate faster with a larger path loss exponent, and thus the inter-cell interference becomes weaker, which motivates a denser deployment of base stations in the network with higher path loss.

\begin{table*}
\begin{center}
\caption{Achievable Rate with Different BS Densities}\label{table:rate1}
\begin{tabular}{ c | c | c |c |c|c|c}
\hline
    \mbox{Carrier frequency} & \mbox{28 GHz}&\mbox{ 28 GHz} &\mbox{28 GHz}&\mbox{28 GHz} &\mbox{2 GHz} &\mbox{2 GHz}\\ \hline
    \mbox{Base station density} & \mbox{Ultra dense}&\mbox{ Dense} &\mbox{Intermediate}&\mbox{Sparse} &- &-\\ \hline

    \mbox{Relative density $\rho$} & 16&4 &1&0.45&-& -\\ \hline
    \mbox{Spectrum efficiency (bps/Hz)} &5.5 & 5.8&4.3&2.7 &4.6&4.6\\ \hline
\mbox{Signal bandwith (MHz)} &100& 100& 100&100&20&100\\ \hline
  \mbox{Achievable rate (Mpbs)} &550 &580&430 & 270&92&459\\ \hline
    \end{tabular}
  \end{center}
\end{table*}

Finally, we compare the spectral efficiency and average achievable rates as a function of the relative density $\rho$ in Table \ref{table:rate1}. We find with a reasonable amount of density, e.g. when the relative density $\rho$ is approximately 1, the \ac{mmWave} system can provide comparable spectrum efficiency as the conventional system at UHF frequencies. With high density, rates that can be achieved are an order of magnitude better than that in the conventional networks, due to the favourable SINR distribution and larger available bandwidth at mmWave frequencies.
\subsection{Comparison with Real-scenario Simulations}
\begin{figure}[!ht]
\centering
\subfigure[center][{Snapshot of the simulated area from Google map.}]{
\includegraphics[width=0.75\columnwidth]{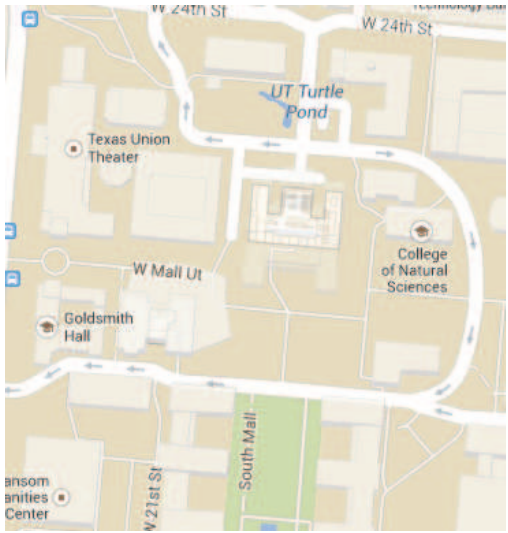}}
\subfigure[center][{{Comparison of SINR distribution}}]{
\includegraphics[width=1\columnwidth]{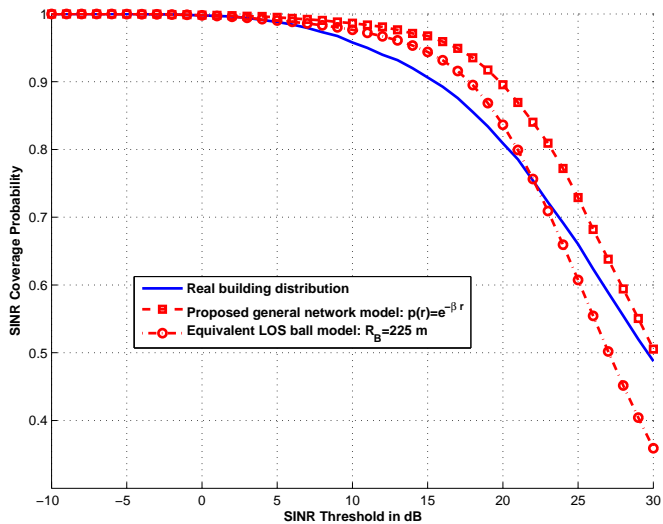}}
\caption{Comparison of SINR coverage results with real-scenario simulations. The snapshot of The University of Texas at Austin campus is from Google map. We use the actual building distribution of the area in the real-scenario simulation. In the simulations of our proposed analytical models, we let $\beta=0.0063\mbox{ m}^{-1}$ in the LOS probability function $p(r)=e^{-\beta r}$, and $R_\mathrm{B}=225$ m in the simplified equivalent LOS ball model, to match the building statistics in the area \cite{bai2013}.}\label{fig:real_simu}
\end{figure}

Now we compare our proposed network models with the simulations using real data. In the real-scenario simulations, we use the building distribution on the campus at The University of Texas at Austin. We also apply a modified version of the base station antenna pattern in \cite{3GPPTR36.9422012} with a smaller beam width of $30^\circ$. The directivity gain at the base station is $M_\mathrm{t}=20$ dB. The mobile station is assumed to use uniform linear array with 4 antennas. When applying our analytical models, we fit the parameters of the LOS probability functions to match the building statistics \cite{bai2013}, and use the sectored model for beamforming pattern. We also assume the mmWave base stations are distributed as a PPP with $r_c=50$ m. As shown in \figref{fig:real_simu}, though some deviations in the high SINR regime, our analytical models generally show a close characterization of the reality. The deviation is explained as follows: the proposed analytical model computes the aggregated SINR coverage probability, averaging over all realizations of building distributions over the infinite plane, while the real-scenario curve only considers a specific realization of buildings in a finite snapshot window. In this case, our model overestimates the coverage probability in the low SINR regime, and underestimates in the high SINR regime, as both signals and interference become more likely to be blocked in the real scenario simulation. We have found in other simulation examples that the reverse can also be true. Our model should be viewed as a characterization of the average distribution and does not necessarily lower or upper bound the distribution for a given realization.


\section{Conclusions}\label{sec:conclusion}
In this paper, we proposed a stochastic geometry framework to analyze coverage and rate in \ac{mmWave} networks for outdoor users and outdoor infrastructure. Our model took blockage effects into account by applying a distance-dependent \ac{LOS} probability function, and modeling the base stations as independent inhomogeneous \ac{LOS} and \ac{NLOS} point processes. Based on the proposed framework, we derived expressions for the downlink SINR and rate coverage probability in \ac{mmWave} cellular networks, which were shown to be efficient in computation and also a good fit with the simulations. We further simplified the blockage model by approximating the random LOS region as a fixed-size equivalent LOS ball. Applying the simplified framework, we analyzed the performance and asymptotic trends in dense networks.


We used numerical results to draw several important conclusions about coverage and rate in mmWave networks.
\begin{itemize}
\item SINR coverage can be comparable to conventional networks at UHF frequency when the base station density is sufficiently high.
\item Achievable rates can be significantly higher than in conventional networks, thanks to the larger available bandwidth.
\item The SINR and rate performance is largely determined by the relative base station density, which is the ratio of the base station density to the blockage density.
\item A transition from a power-limited regime to an interference-limited regime is also observed in mmWave networks, when increasing base station density.
\item The optimal SINR and rate coverage can be achieved with a finite base station density; as increasing base station density need not improve SINR in a (ultra) dense mmWave network.
\end{itemize}


In future work, it would be interesting to analyze the networks with overlaid microwave macrocells, as mmWave systems will co-exist with base stations operated in UHF bands. It would be another interesting topic to incorporate mmWave hardware constraints in the system analysis, and investigate the performance of mmWave networks applying analog/ hybrid beamforming \cite{Ayach2013} or using low-resolution A/D converters at the receivers \cite{mo2014}.
\section*{Acknowledgement}
The authors would like to thank Dr. Xinchen Zhang for his valuable feedback on early drafts of this paper.

\section*{Appendix A}
We provide two useful inequalities in the following lemmas. The first lemma approximates the tail probability of a gamma random variable.
\begin{lemma}[From \cite{alzer1997}]
\label{lem:inequ1} Let $g$ be a normalized gamma random variable with parameter $N$. For a constant $\gamma>0$, the probability $\mathbb{P}(g<\gamma)$ can be tightly upper bounded by
$$
\mathbb{P}(g<\gamma)<\left[1-\mathrm{e}^{-a \gamma}\right]^N,
$$
where $a=N(N!)^{-\frac{1}{N}}$.
\end{lemma}

The following inequality will be used in the dense network analysis.
\begin{lemma}[From \cite{alzer1997}]\label{lem:inequ2}
For $x>0$, it holds that
$$
-\log(1-\mathrm{e}^{-a x})\le\int_{x}^{\infty}\frac{\mathrm{e}^{-t}}{t}\mathrm{d}t\le-\log(1-\mathrm{e}^{-bx}),
$$
where $a=\mathrm{e}^{0.5772}$ and $b=1$. Further, the lower bound generally provides a close approximation.
\end{lemma}
\section*{Appendix B}
{\noindent\it Proof of Lemma \ref{lem:prob}}: For $s=\{\mathrm{L},\mathrm{N}\}$, let $d_s$ be the distance from the typical user to its nearest base station in $\Phi_s$. Note that it is possible that the user observes no base stations in $\Phi_s$. The user is associated with a base station in $\Phi_\mathrm{L}$ if and only if it has a LOS base station, and its nearest base station in $\Phi_\mathrm{L}$ has smaller path loss than that of the nearest base station in $\Phi_\mathrm{N}$. Hence, it follows that
\begin{align}
A_\mathrm{L}&=B_\mathrm{L}\mathbb{P}\left(C_\mathrm{L} d_\mathrm{L}^{-\alpha_\mathrm{L}}>C_\mathrm{N} d_\mathrm{N}^{-\alpha_\mathrm{N}}\right)\nonumber\\
&\stackrel{(a)}=B_\mathrm{L}\int_{0}^{\infty}\mathbb{P}\left(d_\mathrm{N}>\psi_\mathrm{L}(x)\right)f_\mathrm{L}(x)\mathrm{d}x,\label{eqn:appendixa1}
\end{align}
\noindent where $B_\mathrm{L}$ is the probability that the user has at least one LOS base stations, (a) follows that by Lemma \ref{lem:distance}, and $f_\mathrm{L}(x)$ is the probability density function of $d_\mathrm{L}$. Next, note that
\begin{align}
\mathbb{P}\left(d_\mathrm{N}>\psi(x)\right)&=\mathbb{P}\left(\Phi_\mathrm{N}\cap\mathcal{B}(0,\psi_\mathrm{L}(x))=\emptyset\right)\nonumber\\&=\mathrm{e}^{-2\pi\lambda\int_{0}^{\psi_\mathrm{L}(x)}(1-p(t))t\mathrm{d}t}, \label{eqn:appendixa2}
\end{align}
where $\mathcal{B}(0,x)$ denotes the ball centered at the origin of radius $x$. Substituting (\ref{eqn:appendixa2}) for (\ref{eqn:appendixa1}) gives Lemma \ref{lem:prob}. 

\section*{Appendix C}
{\noindent\it Proof of Theorem \ref{thm:coverage}}: Given that the user is associated with a base station in $\Phi_\mathrm{L}$, by Slivnyak's Theorem \cite{BacBla:Stochastic-Geometry-and-Wireless:09}, the conditional coverage probability can be computed as
\begin{align}
P_{\mathrm{c},\mathrm{L}}(T)&=\int_{0}^{\infty}\mathbb{P}\left[C_\mathrm{L}h_0M_\mathrm{r}M_\mathrm{t}x^{-\alpha_\mathrm{L}}>T\left(\sigma^2+I_\mathrm{L}+I_\mathrm{N}\right)\right]\nonumber\\&\hat{f}_\mathrm{L}(x)\mathrm{d}x,
\end{align}
where $I_\mathrm{L}=\sum_{\ell:X_\ell\in\Phi_\mathrm{L}\cap\bar{\mathcal{B}}(0,x)}C_\mathrm{L}\left|h_\ell\right|^2 D_\ell R_\ell^{-\alpha_\mathrm{L}}$ and $I_\mathrm{N}=\sum_{\ell:X_\ell\in\Phi_\mathrm{N}\cap\bar{\mathcal{B}}(0,\psi_\mathrm{L}(x))}C_\mathrm{N}\left|h_\ell\right|^2 D_\ell R_\ell^{-\alpha_\mathrm{N}}$ are the interference strength from the tiers of LOS and NLOS base stations, respectively. Next, noting that $\left|h_0\right|^2$ is a normalized gamma random variable with parameter $N_\mathrm{L}$, we have the following approximation
\begin{align}
&\mathbb{P}\left[C_\mathrm{L}h_0M_\mathrm{r}M_\mathrm{t}x^{-\alpha_\mathrm{L}}>T\left(\sigma^2+I_\mathrm{L}+I_\mathrm{N}\right)\right]\nonumber\\
&=\mathbb{P}\left[h_0>x^{\alpha_\mathrm{L}}T\left(\sigma^2+I_\mathrm{L}+I_\mathrm{N}\right)/(C_\mathrm{L}M_\mathrm{r}M_\mathrm{t})\right]\nonumber\\
&\stackrel{(a)}\approx 1-\mathbb E_\Phi\left[\left(1-\mathrm{e}^{-\frac{\eta_\mathrm{L} x^{\alpha_\mathrm{L}}T\left(\sigma^2+I_\mathrm{L}+I_\mathrm{N}\right)}{C_\mathrm{L}M_\mathrm{r}M_\mathrm{t}}}\right)^{N_\mathrm{L}}\right]\nonumber\\
&\stackrel{(b)}=\sum_{n=1}^{N_\mathrm{L}}(-1)^{n+1}{N_\mathrm{L}\choose{n}}\mathbb E_{\Phi}\left[\mathrm{e}^{-\frac{n \eta_\mathrm{L} x^{\alpha_\mathrm{L}}T\left(\sigma^2+I_\mathrm{L}+I_\mathrm{N}\right)}{C_\mathrm{L}M_\mathrm{r}M_\mathrm{t}}}\right]\nonumber\\
&\stackrel{(c)}=\sum_{n=1}^{N_\mathrm{L}}(-1)^{n+1}{N_\mathrm{L}\choose{n}}\mathrm{e}^{-\frac{n \eta_\mathrm{L} x^{\alpha_\mathrm{L}}T\sigma^2}{C_\mathrm{L}M_\mathrm{r}M_\mathrm{t}}}\mathbb{E}_{\Phi_\mathrm{L}}\left[\mathrm{e}^{-\frac{n \eta_\mathrm{L} x^{\alpha_\mathrm{L}}TI_\mathrm{L}}{C_\mathrm{L}M_\mathrm{r}M_\mathrm{t}}}\right]\nonumber\\&\mathbb{E}_{\Phi_\mathrm{N}}\left[\mathrm{e}^{-\frac{n \eta_\mathrm{L} x^{\alpha_\mathrm{L}}TI_\mathrm{N}}{C_\mathrm{L}M_\mathrm{r}M_\mathrm{t}}}\right]\label{eqn:thm2},
\end{align}
where $\eta_\mathrm{L}=N_\mathrm{L}(N_\mathrm{L}!)^{-\frac{1}{N_\mathrm{L}}}$, $(a)$ is from Lemma \ref{lem:inequ1} \cite{alzer1997} in Appendix A, $(b)$ follows from Binomial theorem and the assumption that $N_\mathrm{L}$ is an integer, and $(c)$ follows from the fact that $\Phi_\mathrm{L}$ and $\Phi_\mathrm{N}$ are independent. Now we apply concepts from stochastic geometry to compute the term for LOS interfering links $\mathbb{E}_{\Phi_\mathrm{L}}\left[\mathrm{e}^{-\frac{n \eta_\mathrm{L} x^{\alpha_\mathrm{L}}TI_\mathrm{L}}{C_\mathrm{L}M_\mathrm{r}M_\mathrm{t}}}\right]$ in (\ref{eqn:thm2}) as
\begin{align*}
\mathbb{E}_{\Phi_\mathrm{L}}&\left[\mathrm{e}^{-\frac{n \eta_\mathrm{L} x^{\alpha_\mathrm{L}}TI_\mathrm{L}}{C_\mathrm{L}M_\mathrm{r}M_\mathrm{t}}}\right]\\&=\mathbb{E}\left[\mathrm{e}^{-\frac{n \eta_\mathrm{L} x^{\alpha_\mathrm{L}}T \sum_{\ell:X_\ell\in\Phi_\mathrm{L}\cap\bar{\mathcal{B}}(0,x)}\left|h_\ell\right|^2 D_\ell R_\ell^{-\alpha_\mathrm{L}}}{M_\mathrm{r}M_\mathrm{t}}}\right]\\
&\stackrel{(c)}=\mathrm{e}^{\left(-2\pi\lambda\sum_{k=1}^{4}b_k\int_{x}^{\infty}\left(1-\mathbb {E}_g\left[\mathrm{e}^{-nT\eta_\mathrm{L} g \bar{a}_k (x/t)^{\alpha_\mathrm{L}}}\right]\right)p(t)t\mathrm{d}t\right)}\\
&\stackrel{(d)}=\prod_{k=1}^{4}\mathrm{e}^{-2\pi\lambda b_k\int_{x}^{\infty}\left(1-1/{\left(1+\eta_\mathrm{L} \bar{a}_k nT(x/t)^{\alpha_\mathrm{L}}/N_\mathrm{L}\right)^{N_\mathrm{L}}}\right)p(t)t\mathrm{d}t}
\\&=\mathrm{e}^{-Q_n(T,x)},
\end{align*}
where $g$ in $(c)$ is a normalized gamma random variable with parameter $N_\mathrm{L}$, $\bar{a}_k=\frac{a_k}{M_\mathrm{t}M_\mathrm{r}}$, and for $1\le k\le4$, $a_k$ and $b_k$ are defined previously in Table \ref{table:constant}; (c) is from computing the Laplace functional of the PPP $\Phi_\mathrm{L}$ \cite{BacBla:Stochastic-Geometry-and-Wireless:09}; $(d)$ is by computing the moment generating function of a gamma random variable $g$. 

Similarly, for the NLOS interfering links, the small-scale fading term $|h_\ell|^2$ is a normalized gamma variable with parameter $N_\mathrm{N}$. Thus, we can compute $\mathbb E_{\Phi_\mathrm{N}}\left[\mathrm{e}^{-\frac{n \eta_\mathrm{L} x^{\alpha_\mathrm{L}}T I_\mathrm{N}}{C_\mathrm{L} M_\mathrm{r}M_\mathrm{t}}}\right]$ as
\begin{align*}
&\mathbb E_{\Phi_\mathrm{N}}\left[\mathrm{e}^{-\frac{n \eta_\mathrm{L} x^{\alpha_\mathrm{L}}T I_\mathrm{N}}{C_\mathrm{L} M_\mathrm{r}M_\mathrm{t}}}\right]\\&=\mathbb{E}\left[\mathrm{e}^{-\frac{n \eta_\mathrm{L} x^{\alpha_\mathrm{L}}TC_\mathrm{N} \sum_{\ell:X_\ell\in\Phi_\mathrm{N}\cap\bar{\mathcal{B}}(0,\psi(x))}\left|h_\ell\right|^2 D_\ell R_\ell^{-\alpha_\mathrm{N}}}{C_\mathrm{L}M_\mathrm{r}M_\mathrm{t}}}\right]\\
&=\prod_{k=1}^{4}\mathrm{e}^{-2\pi\lambda b_k\int_{\psi_\mathrm{L}(x)}^{\infty}\left(1-1/{\left(1+\frac{\eta_\mathrm{L} \bar{a}_k nTC_\mathrm{N}x^{\alpha_\mathrm{L}}}{C_\mathrm{L} t^{\alpha_\mathrm{N}}N_\mathrm{N}}\right)^{N_\mathrm{N}}}\right)\left(1-p(t)\right)t\mathrm{d}t}
\\&=\mathrm{e}^{-V_n(T,x)}.
\end{align*}
Then, we obtain (\ref{eqn:pc1}) from (\ref{eqn:thm2}) by the linearity of integrals.

Given the user is associated with a \ac{NLOS} base station, we can also derive the conditional coverage probability $P_{c,\mathrm{N}}(T)$ following same approach as that of $P_{c,\mathrm{L}}(T).$ Thus, we omit the detailed proof of (\ref{eqn:pc2}) here.

Finally, by the law of total probability, it follows that
$P_\mathrm{c}(T)=A_\mathrm{L}P_{\mathrm{c},\mathrm{L}}(T)+A_\mathrm{N}P_{\mathrm{c},\mathrm{N}}(T).$ 

\section*{Appendix D}
{\noindent\it Proof Sketch of Theorem \ref{thm:dense}}: For a general $\alpha_\mathrm{L}$, the coverage probability can be computed as
\begin{align*}
P_\mathrm{c}(T)&=A_\mathrm{L}P_{\mathrm{c},\mathrm{L}}(T)=A_\mathrm{L}\mathbb P(\mbox{SIR}>T)
\\&=A_\mathrm{L}\int_{0}^{R_\mathrm{B}}\mathbb{P}(C_\mathrm{L}M_rM_tr^{-\alpha_\mathrm{L}}>TI_r)\frac{2\pi\lambda r}{A_\mathrm{L}}\mathrm{e}^{-\lambda\pi r^2}\mathrm{d}r,
\end{align*}
where $I_r=\sum_{X_\ell\in\Phi\cap\left(\mathcal{B}(0,R_\mathrm{B})\slash\mathcal{B}(0,R_\mathrm{B})\right)}D_\ell C_\mathrm{L} R_\ell^{-\alpha_\mathrm{L}}$ is the interference power given that the distance to the user's serving base station is $R_0=r$. Next, the probability $\mathbb{P}(C_\mathrm{L}M_rM_tr^{-\alpha_\mathrm{L}}>TI_r)$ can be approximated as
\begin{align}
&\mathbb{P}(C_\mathrm{L}M_rM_tr^{-\alpha_\mathrm{L}}>TI_r)\stackrel{(a)}\approx\mathbb{P}(g>Tr^{\alpha_\mathrm{L}}I_r/(C_\mathrm{L}M_rM_t))\nonumber\\
&\stackrel{(b)}\approx1-\mathbb E_{\Phi_\mathrm{L}}\left[\left(1-\mathrm{e}^{-\eta Tr^{\alpha_\mathrm{L}}I_r/(C_\mathrm{L}M_rM_t)}\right)^N\right]\nonumber\\
&=\sum_{\ell=1}^{N}{N\choose{\ell}}(-1)^\ell \mathbb{E}_{\Phi_\mathrm{L}}\left[\mathrm{e}^{-\ell\eta Tr^{\alpha_\mathrm{L}}I_r/(C_\mathrm{L}M_rM_t)}\right].\label{eqn:Dpc1}
\end{align}
In $(a)$, the dummy variable $g$ is a normalized gamma variable with parameter $N$, and the approximation in $(a)$ follows from the fact that a normalized Gamma distribution converges to identity when its parameter goes to infinity, i.e., $\lim_{n\to\infty}\frac{n^nx^{n-1}\mathrm{e}^{-nx}}{\Gamma(n)}=\delta(x-1)$ \cite{Aris1999}, where $\delta(x)$ is the Dirac delta function. In $(b)$, it directly follows from Lemma \ref{lem:inequ1} by taking $\eta=N(N!)^{1/N}$.

Next, we can compute $\mathbb{E}_{\Phi_\mathrm{L}}\left[\mathrm{e}^{-\ell\eta Tr^{\alpha_\mathrm{L}}I_r/(C_\mathrm{L}M_rM_t)}\right]$ as
\begin{align}
\mathbb{E}_{\Phi_\mathrm{L}}&\left[\mathrm{e}^{-\ell\eta Tr^{\alpha_\mathrm{L}}I_r/(C_\mathrm{L}M_rM_t)}\right]\nonumber\\\stackrel{(c)}=&\exp\left(\sum_{k=1}^{4}-2\pi\lambda b_k\int_{r}^{R_\mathrm{B}}1-\mathrm{e}^{-\ell\eta \bar{a}_k T(r/t)^{\alpha_\mathrm{L}}}t\mathrm{d}t\right)\nonumber\\
\stackrel{(d)}=&\mathrm{e}^{-\pi\lambda(R_\mathrm B^2-r^2)}\times\nonumber\\&\mathrm{e}^{\sum_{k=1}^{4}\frac{2}{\alpha_\mathrm{L}}\pi\lambda r^2 b_k(T\ell\eta\bar{a}_k)^{2/\alpha_\mathrm{L}}\int_{\ell\eta T\bar{a}_k(r/R_\mathrm{B})^{\alpha_\mathrm{L}}}^{\ell\eta T \bar{a}_k}\frac{\mathrm{e}^{-s}}{s^{1+2/\alpha_\mathrm{L}}}\mathrm{d}s}\label{eqn:Dpc3}\\
=&\mathrm{e}^{-\pi\lambda(R_\mathrm B^2-r^2)}\times\nonumber\\&\mathrm{e}^{\sum_{k=1}^{4}\frac{2}{\alpha_\mathrm{L}}\pi\lambda r^2b_k(T\ell\eta\bar{a}_k)^{2/\alpha_\mathrm{L}}\Gamma\left(\frac{-2}{\alpha_\mathrm{L}};\ell\eta T\bar{a}_k(r/R_\mathrm{B})^{\alpha_\mathrm{L}},\ell\eta T \bar{a}_k\right)},\label{eqn:Dpc2}
\end{align}
where (c) follows from computing the Laplace functional of the PPP $\Phi_\mathrm{L}$ \cite{BacBla:Stochastic-Geometry-and-Wireless:09}, and (d) follows from changing variable as $s=\ell\eta \bar{a}_k T(r/t)^{\alpha_\mathrm{L}}$. Hence, (\ref{eqn:thmdense1}) directly follows from substituting (\ref{eqn:Dpc2}) for (\ref{eqn:Dpc1}) and letting $\rho=\pi\lambda R_\mathrm{B}^2$.

When $\alpha_\mathrm{L}=2$, the steps above hold true till (\ref{eqn:Dpc3}), which can be further simplified as
\begin{align}
&\mathbb{E}_{\Phi_\mathrm{L}}\left[\mathrm{e}^{-\ell\eta Tr^{\alpha_\mathrm{L}}I_r/(C_\mathrm{L}M_rM_t)}\right]\nonumber\\&=\mathrm{e}^{-\pi\lambda(R_\mathrm B^2-r^2)}\mathrm{e}^{\sum_{k=1}^{4}\left(\pi\lambda r^2 b_kT\ell\eta\bar{a}_k\int_{\ell\eta T\bar{a}_k(r/R_\mathrm{B})^{2}}^{\ell\eta T \bar{a}_k}\frac{\mathrm{e}^{-s}}{s^{2}}\mathrm{d}s\right)}\nonumber\\
&\stackrel{(e)}=\mathrm{e}^{-\pi\lambda (R_\mathrm B^2-r^2)}\times\nonumber\\&\mathrm{e}^{\sum_{k=1}^{4}\pi\lambda r^2 b_k\left(\frac{\mathrm{e}^{-(r/R_\mathrm{B})^2\ell T\ell\eta\bar{a}_k}}{(r/R_\mathrm{B})^2}-\mathrm{e}^{-\ell T\ell\eta\bar{a}_k}+\int_{\ell T\ell\eta\bar{a}_k (r/R_\mathrm{B})^2}^{\ell T\ell\eta\bar{a}_k}\frac{\mathrm{e}^{-s}}{s}\mathrm{d}s\right)}\nonumber\\
&\stackrel{(f)}\approx\mathrm{e}^{-\pi\lambda (R_\mathrm B^2-r^2)}\exp\left(\sum_{k=1}^{4}\pi\lambda r^2 b_k\left(\frac{\mathrm{e}^{-(r/R_\mathrm{B})^2\ell T\ell\eta\bar{a}_k}}{(r/R_\mathrm{B})^2}\right.\right.\nonumber\\&\left.\left.-\mathrm{e}^{-\ell T\ell\eta\bar{a}_k}-\log\frac{1-\mathrm{e}^{\mu\ell T\ell\eta\bar{a}_k (r/R_\mathrm{B})^2}}{1-\mathrm{e}^{\mu\ell T\ell\eta\bar{a}_k}}\right)\right),\label{eqn:Dpc4}
\end{align}
where $(e)$ is from computing integration by part, (f) follows from Lemma \ref{lem:inequ2} by letting $\mu=\mathrm{e}^{0.5772}$. Lastly, (\ref{eqn:thmdense2}) follows from substituting (\ref{eqn:Dpc4}) for (\ref{eqn:Dpc1}) and letting $\rho=\pi\lambda R_\mathrm{B}^2$. 

\section*{Appendix E}
{\noindent \it Proof Sketch of Theorem \ref{thm:asymp}}: First, we show that the exact SIR distribution in our dense network model is unchanged when the relative density $\rho=\lambda \pi R_\mathrm{B}^2$ is fixed. Let $F_\ell=R_\ell^{\alpha_\mathrm{L}}$ be the path loss gain in the $\ell$-th link. Then the SIR expression in (\ref{eqn:SIR}) can be rewritten as $\mbox{SIR}=M_\mathrm{t}M_\mathrm{r} F_0^{-1}/\left(\sum_{\ell:X_\ell\in\Phi_\mathrm{L}}D_\ell F_\ell^{-1}\right)$, where $D_\ell$ is the directivity gain in the $\ell$-th link. Further, using displacement theorem \cite{BacBla:Stochastic-Geometry-and-Wireless:09} and the method in \cite[Lemma 7]{bai2013b}, we can show that $\{F_\ell\}_{\ell:X_\ell\in\Phi_\mathrm{L}}$ forms a PPP on the interval $\left(0, R_\mathrm{B}^{\alpha_\mathrm{L}}\right)$ with density measure function $\Lambda(0,t)=\lambda \pi t^{2/{\alpha_\mathrm{L}}}$. Next, for $c>0$, we define an class of {\it equivalent networks} as the networks with base station density $c\lambda$ and LOS range $c^{-0.5}R_B $. Note that all networks in the equivalent class have the same relative density $\rho=\lambda \pi R_\mathrm{B}^2$ as the original network. Then we can show that a scaled version of path loss gain process $\left\{c^{\alpha_\mathrm{L}/2} F_\ell\right\}_\ell$ in an equivalent network has the exact same distribution, more specifically, the measure density function, as $\{F_\ell\}_\ell$ in the original network. Thus, all equivalent networks have the same SIR distribution as the original network; as the scaling constant $c^{\alpha_\mathrm{L}/2}$ cancels in the SIR expression. So far, we have shown that the SIR distribution in dense networks is unchanged when the relative density $\rho$ is fixed.

Given the SIR invariant property with respect to $\rho$, one way to investigate the asymptotic SIR when $\rho\to\infty$ is to fix $\lambda$ as a constant $\lambda_0$ and examine the SIR performance when $R_\mathrm{B}\to\infty$. In other words, when $\rho\to\infty$, the asymptotic SIR in the original network has the same distribution as the SIR in its {\it asymptotic equivalent network}, which has a base station density of $\lambda_0$, and an infinitely large $R_\mathrm{B}$. Let $\mbox{SIR}_0$ be the SIR in the asymptotic equivalent network. Note that in the asymptotic equivalent network, \ac{LOS} base stations form a homogeneous PPP with density $\lambda_0$ on the entire plane.

Next, for any realization of the path loss process $\{F_\ell\}_{\ell:X_\ell\in\Phi_\mathrm{L}}$, the SIR in the asymptotic equivalent network can be lower and upper bounded by assuming all interfering links achieve the maximum directivity gain $M_\mathrm{r}M_\mathrm{t}$ and the minimum directivity gain $m_\mathrm{r}m_\mathrm{t}$, respectively, as
\begin{align}
\frac{F_0^{-1}}{\sum_{\ell:X_\ell\in\Phi_\mathrm{L}} F_\ell^{-1}}\le\mbox{SIR}_0\le\frac{M_\mathrm{r}M_\mathrm{t} F_0^{-1}}{m_\mathrm{r}m_\mathrm{t}\sum_{\ell:X_\ell\in\Phi_\mathrm{L}} F_\ell^{-1}}.\label{eqn:AppendixE}
\end{align}

For $\alpha_\mathrm{L}>2$, an lower bound of the SIR coverage probability in the asymptotic equivalent network can be computed as for $T>1$,
\begin{align*}
\mathbb{P}(\mbox{SIR}_0>T)&\ge\mathbb{P}\left(\frac{F_0^{-1}}{\sum_{\ell:X_\ell\in\Phi_\mathrm{L}} F_\ell^{-1}}>T\right)\nonumber\\&\stackrel{(a)}=\frac{\alpha_\mathrm{L} T^{-2/\alpha_\mathrm{L}}}{2\pi\sin(2\pi/\alpha_\mathrm{L})},
\end{align*}
where $(a)$ follows from Proposition 10 in \cite{Blaszczyszyn2013}.

Finally, for $\alpha_\mathrm{L}\le2$, we show that the upper bound of $\mbox{SIR}_0$ in (\ref{eqn:AppendixE}) converges to zero in probability. By \cite[Proposition 1, Fact 1]{zhang2013}, the distribution of the upper bound expression in (\ref{eqn:AppendixE}) is invariant with the small-scale fading distribution in the asymptotic equivalent network, and thus has the same distribution as in the {\it Rayleigh-fading network} investigated in \cite{Dhillon2012}, where independent Rayleigh fading is assumed in each links. Consequently, we compute the coverage probability for the $\mbox{SIR}_0$ upper bound in (\ref{eqn:AppendixE}) as follows: for all $T>0$,
\begin{align*}
&\mathbb{P}\left(\mbox{SIR}_0>T\right)\stackrel{(b)}<\mathbb{P}\left[\frac{M_\mathrm{r}M_\mathrm{t} F_0^{-1}}{m_\mathrm{r}m_\mathrm{t}\sum_{\ell>0:X_\ell\in\Phi_\mathrm{L}} F_{\ell}^{-1}}>T\right]\\
&=\mathbb{P}\left[\bigcup_{\ell:X_\ell\in\Phi_\mathrm{L}}\frac{M_\mathrm{r}M_\mathrm{t} F_\ell^{-1}}{m_\mathrm{r}m_\mathrm{t}\sum_{\ell'\ne\ell:X_\ell'\in\Phi_\mathrm{L}} F_{\ell'}^{-1}}>T\right]\\
&=\mathbb{P}\left[\bigcup_{\ell:X_\ell\in\Phi_\mathrm{L}}\frac{F_\ell^{-1}}{m_\mathrm{r}m_\mathrm{t}\sum_{\ell'\ne\ell:X_\ell'\in\Phi_\mathrm{L}} F_{\ell'}^{-1}}>\frac{m_\mathrm{r}m_\mathrm{t}T}{M_\mathrm{r}M_\mathrm{t} }\right]\\
&\stackrel{(c)}\le\sum_{X_\ell\in\Phi_\mathrm{L}}\mathbb{P}\left[\frac{F_\ell^{-1}}{m_\mathrm{r}m_\mathrm{t}\sum_{\ell'\ne\ell:X_\ell'\in\Phi_\mathrm{L}} F_{\ell'}^{-1}}>\frac{m_\mathrm{r}m_\mathrm{t}T}{M_\mathrm{r}M_\mathrm{t} }\right]\\
&\stackrel{(d)}=\pi\lambda\int_{0}^{\infty}\exp\left(-\pi\lambda \bar{T}^{2/\alpha_{\mathrm{L}}} t \int_{0}^{\infty}\frac{1}{1+r^{\alpha_{\mathrm{L}}/2}} \mathrm{d}r\right)\mathrm{d}t\\
&=\frac{\bar{T}^{-2/\alpha_{\mathrm{L}}}}{\int_{0}^{\infty}\frac{1}{1+r^{\alpha_{\mathrm{L}}/2}} \mathrm{d}r}\\&\stackrel{(e)}=0,
\end{align*}
where $\bar{T}=\frac{m_\mathrm{r}m_\mathrm{t}T}{M_\mathrm{r}M_\mathrm{t} }$, $(b)$ is obtained from (\ref{eqn:AppendixE}), (c) is from the union bound, (d) follows from the fact that the upper bound in (\ref{eqn:AppendixE}) has the same distribution as in the Rayleigh-fading network in \cite{Dhillon2012}, and the similar algebra in \cite[Appendix B, Eqn (15)-(17)]{Dhillon2012}, and (e) is from the fact that $\int_{0}^{\infty}\frac{1}{1+r^{\alpha_{\mathrm{L}}/2}} \mathrm{d}r$ is infinity when $\alpha_\mathrm{L}\le2$. By now, we have shown that an upper bound of SIR in the asymptotic equivalent network, which also upper bounds the SIR in the original network when $\rho\to\infty$, converges to zero in probability.\endproof

\bibliographystyle{IEEEtran}


\end{document}